\documentclass[aps]{revtex4}

\usepackage{amsmath,amsfonts}
\usepackage[dvips]{graphicx}
 
\newcommand{\tr}{\mathop{\text{tr}}}

\begin{document}
 
\title{Dynamics of the Universe with global rotation}
 
\author{W{\l}odzimierz God{\l}owski}
\email{godlows@oa.uj.edu.pl}
 
\author{Marek Szyd{\l}owski}
\email{uoszydlo@cyf-kr.edu.pl}
 
\affiliation{Astronomical Observatory, Jagiellonian University, \\
Orla 171, 30-244 Krakow, Poland}
 
\begin{abstract}
We analyze dynamics of the FRW models with global rotation in terms of
dynamical system methods. We reduce dynamics of these models to the FRW 
models with some fictitious fluid which scales like radiation matter. 
This fluid mimics dynamically effects of global rotation.
The significance of the global rotation of the Universe for the resolution of the
acceleration and horizon problems in cosmology is investigated. It is found
that dynamics of the Universe can be reduced to the two-dimensional Hamiltonian
dynamical system. Then the construction of the Hamiltonian allows for full
classification of evolution paths. On the phase portraits we find the domains 
of cosmic acceleration for the globally rotating universe as well as the 
trajectories for which the horizon problem is solved. We show that the FRW 
models with global rotation are structurally stable. This proves that 
the universe acceleration is due to the global rotation. 
It is also shown how global rotation gives a natural explanation of the
empirical relation between angular momentum for clusters and superclusters
of galaxies. The relation $J \sim M^2$ is obtained as a consequence of self
similarity invariance of the dynamics of the FRW model with global rotation.
In derivation of this relation we use the Lie group of symmetry analysis of 
differential equation.


\end{abstract}

\maketitle

\section{Introduction}
 
Rotation is a very universal phenomenon in nature. We can observe rotating 
objects at all scales of the Universe, from the elementary particles 
to planets, stars and galaxies. The question is, whether this property is an 
attribute of the whole universe at a very large scale structure. On the other 
hand, if our universe does not rotate, then the question is why and how does 
this happen? Since  rotation is generic in the universe, the possible rotation 
of the universe cannot be excluded at the very beginning. 
Moreover, we should explain the physical mechanism which prevents 
universal rotation of the universe \cite{Obukhov02}. 

Of course, if the Universe rotate, then there should exist some observational 
manifestations of this global rotation. Following the standard approach 
we can study the motion of test particles and photons on the background 
of corresponding spacetimes which admit the rotation. Our idea is to test 
the effects of rotation by classical cosmological tests like luminosity distance 
or angular size of radiogalaxies.
 
Since the work of Lanczos \cite{Lanczos24}, Gamov \cite{Gamov46}, Goedel
\cite{Goedel49}, and Hawking \cite{Hawking69}, the cosmological models with 
rotation have been studied as well as the behavior of geodesics in such 
spacetimes. Although the quite strong upper limits for cosmic vorticity were 
obtained from analysis of CMB or BBN 
\cite{Hawking69,Collins73,Barrow85,Bunn96,Kogut97}
all these works based on the model
in which shear and vorticity effects are inseparable, i.e., in the sense that
zero shear automatically implies zero vorticity. Therefore, the obtained limits
actually placed not on the vorticity, but rather on the shear induced by it
within the specific geometry of spacetime \cite{Obukhov02}. Finally, one thus 
needs a separate analysis of these limits in expanding cosmological models 
with trivial shear (shear-free) but non-zero rotation and expansion (it is 
an idea of Obukhov \cite{Obukhov02}). Principally, there are two classes of 
such models in which this analysis can be performed
\begin{enumerate}
\item
Newtonian shear-free models so-called Heckmann-Sch{\"u}cking models
\cite{Heckmann59,Heckmann61};
\item
general relativistic spatially homogeneous models with the geometry of the
Bianchi IX and the cosmological constant \cite{Obukhov02}.
\end{enumerate}

Let us consider the first class of models.
We have a homogeneous Newtonian universe defined on product of
the three-dimensional Euclidean space and absolute time coordinate.
This universe is homogeneous, density and pressure of the fluid
have no spatial dependence and the velocity vector field depends
linearly on spatial coordinate \cite{Szekeres77,Senovilla98}.
In this case contrary to the general relativity theory there exist
shear-free solutions which satisfy the Heckmann-Sch{\"u}cking equations 
and describe both expanding and rotating universes \cite{Godlowski03}.
The rotating homogeneous non-tilted universes when considered on a
relativistic level filled with perfect fluid must have non-vanishing shear
\cite{King73,Raychaudhuri79}. Therefore, it is very difficult to
consider both effects of anisotropy and rotation, that's why
we study the Newtonian counterpart of model with rotation.
Moreover, it is reasonable to assume that the shear scalar is
sufficiently small compared with angular velocity scalar since
the shear falls off more rapidly than the rotation as the universe
expands \cite{Hawking69,Ellis73,Li98,Barrow04}.

To compare the results of analysis of supernovae data with a general
relativistic model we formally consider a non-flat model $\Omega_{k,0}\ne 0$, 
although the satisfactory interpretation of $\Omega_{k,0}$ can be found 
beyond Newtonian cosmology in the general relativity.

The motion of the fluid in a homogeneous Newtonian universe is
described by the scalar expansion $\theta$, the rotation tensor
$\omega_{ab}$, and the shear tensor $\sigma_{ab}$. The homogeneous
rotation of fluid as a whole is usually called the global rotation of
the universe \cite{Li98}.

The problem of the rotation of the whole Universe attracted the attention
of several scientists. It was shown, that the reported values are too big,
when compared with the CMB anisotropy. Silk \cite{Silk70} pointed out that
presently the dynamical effects of a general rotation of the Universe are
unimportant, contrary to the the early universe, when angular velocity
$\omega_0 \ge 10^{-13}$ rad/yr. He stressed that now the period of rotation
must be greater than the Hubble time, which is consequence of the CMB isotropy. 
Barrow \cite{Barrow85} also addressed this question and showed that
the cosmic vorticity depends strongly on the cosmological models and
assumptions connected with linearisation of homogeneous, anisotropic
cosmological models over the isotropic Friedmann universe. For the flat
universe they obtained the value ($\omega/H_0) = 2 \cdot 10^{-5}$.

While the investigations of the cosmological models with global rotation has
a long history, the global rotation of the Universe is still not detectable 
by observation. Some observational evidences for the global rotation of the 
Universe have not been confirmed \cite{Birch82,Phinney83,Bietenholz84}. 
The cosmological origin of the rotation of astronomical objects was analysed 
by Fil'chenkov in the framework of quantum cosmology and tunnelling approach 
\cite{Filchenkov03}.

The alignment of galaxies along the direction of the global rotation would 
also be some indication of the discussed phenomenon. However, the primordial 
dipole anisotropy of the distribution of rotation axes of galaxies is blurred 
due to the irregular shape of the protogalaxy, which can lead, after phase 
transition, to an almost random distribution. It is clear that the angular 
momentum of a structure is a sum of spins of components and its own angular 
momentum. A lot of work has been done in study of the alignment of galaxies 
in superclusters (for review see \cite{Djorgovski87}). 
Here, the result are ambiguous, but several independent investigations 
claimed the existence of the weak alignment of galaxies in respect to the 
supercluster main plane \cite{Flin86,Kashikawa92,Godlowski93,Godlowski94}.

Independent observations of Supernovae type Ia (SNIa) have indicated that 
our Universe presently accelerates 
\cite{Perlmutter98,Perlmutter99,Garnavich98,Riess98}. There is a fundamental
problem for theoretical physics to explain the origin of this acceleration.
One of the first explanations was the cosmological constant. It is also 
possible to introduce the global rotation as we do in the present paper. 
While such a possibility seems to be attractive in the context of the 
accelerating universe, the cosmological constant is still needed to explain 
SNIa data because the contribution to acceleration coming from the global 
rotation is too small \cite{Godlowski03a}.

In our previous papers we study some constraints on rotation coming from 
SNIa data, primordial nucleosynthesis as well as from the age of the oldest 
in the Universe \cite{Godlowski03a}. 
The main aims of this paper is rather theoretical than empirical. We show 
that the analysis of the model dynamics is more effective after rewriting 
the basic equation to the new form with dimensionless quantities. Then we 
obtain a two-dimensional dynamical system in which coefficients have a simple
interpretation of a density parameter of fluid fulfilling the Universe 
in the present epoch. In this approach dynamics is described by the 
Friedman-Robertson-Walker (FRW) model with some additional noninteracting 
fluid for which we can define density parameter 
$\Omega_{\omega}\equiv \rho_{\omega}/3H_{0}^{2}$.
The presented formalism gives us also a natural base to discuss the influence
of global rotation on acceleration of the Universe. From the observational 
constrains it is possible to  find a limit for the angular velocity of the 
Universe. For example, Ciufolini and Wheeler \cite{Ciufolini95} obtained
$\omega_0 \simeq 6 \cdot 10^{-21} \text{rad s}^{-1} \simeq 2 \cdot 10^{-13}
\text{rad yr}^{-1}$ from the CMB limit for the homogeneous B(IX) models with 
rotation. The limits for angular velocity of the Universe obtained 
from the CMB and SNIa were also discussed \cite{Godlowski03a}.
Here we also investigate some dynamical effects of the global rotation on the
solving classical problems in cosmology, namely the horizon, flatness and 
cosmological constant problems.

Organization of our paper is as follows. In section 2 we present Hamiltonian 
dynamics of the FRW model with global rotation. In section 3 we discuss some 
properties of exact solutions with rotation. In section 4 we discuss the 
domains of the phase plane for which the cosmic acceleration and horizon 
problem are solved. In section 5 we construct the phase portraits for the 
FRW model with rotation and dust ($\gamma = 1$). We also investigate general 
properties of dynamics by showing that generally it can be reduced to the 
form of the FRW models with non-interacting multifluid scaling like radiation. 
In section 6 a detailed analysis of dynamics on phase portrait is given
in two dimensional phase plane. In section 7 we discuss the flatness and 
cosmological constant problems in the context of model with rotation.
In section 8, it is proved (without of any help of empirical evidence) that 
formulae $J \sim M^2$ is exactly valid for the universe with dust 
($\gamma = 1$). For proof we use the Lie group theory of symmetry analysis 
of differential equations, describing the evolution of the universe. Then 
the relation between angular momentum and mass becomes a constraint relation 
between invariants of a self similarity group of symmetry of the basic 
equations. In section 9 we summarise obtained results.

\section{Reduced Hamiltonian Dynamics of FRW Universe with global rotation}

In this paper we adopt the Hamiltonian formalism. It gives immediate insight 
into possible motion because the dynamical problem is analogous to that of 
a particle of unit mass moving in the one-dimension potential.

The Raychaudhuri equation describing the relation among scalar expansion
$\Theta$, the rotation tensor $\omega_{ab}$ and shear tensor $\sigma_{ab}$ 
(if the acceleration vector $A^a \equiv u^b \nabla_v u^a$ vanishes, which 
would necessarily follow in the case of dust) has the form \cite{Ciufolini95}
\begin{equation}
\label{eq:1}
\dot{\Theta} = - \frac{1}{3} \Theta^2 - 2(\sigma^2 - \omega^2)
+ \Lambda - 4 \pi G (\rho +3p)
\end{equation}
where $\omega^2 = \omega_{ab} \omega^{ab}/2$, $\sigma^2=\sigma_{ab}
\sigma^{ab}/2$, $\Lambda$ is the cosmological constant, $G$ is the 
gravitational constant, the fluid
is perfect fluid with the stress-energy tensor
$T_{ab} = (\rho + p) u_a u_b + pg_{ab}$ ($\rho$ is the mass density and
$p$ is its pressure). The shear scalar $\sigma^2$ vanishes for the FRW
model, while for the anisotropic Bianchi I or Bianchi V model (hereafter 
B(I) and B(V)) it satisfies the following equation
\begin{equation}
\label{eq:2}
\frac{d\sigma^2}{dt} = - 2 \Theta \sigma^2.
\end{equation}
Our aim is to consider the FRW model and its generalisation---anisotropic
B(I) or B(V) models with global rotation. In the anisotropic generalization
we have
\[
3\Theta = (H_1 + H_2 + H_3), 
\]
where $H_i = d(\ln a_i)/dt$ are the Hubble functions in three different 
main directions, and $a_i$ ($i=1,2,3$) are corresponding scale factors. 
Equation~(\ref{eq:1}) can be also obtained within the Newtonian theory. 
However, in general relativity for the perfect fluid model with 
vanishing shear $\sigma=0$, and acceleration $\dot{u}_a=0$, 
the field equations can be consistent only if $\omega \Theta=0$. 
For the dust matter fluid with $\sigma=0$, $\dot{u}_a=0$, particles of 
dust are moving along geodesics. This is in contrast to the Newtonian
case \cite{Ellis73}. Therefore, expanding and rotating spatially homogeneous
universes filled with perfect fluid must have a non-interacting shear.
Li \cite{Li98} considered the case that $\sigma$ is sufficiently small 
compared with $\omega$ since the shear following equation (\ref{eq:2}) 
falls off more rapidly than the rotation, during the expansion of the Universe
\cite{Hawking69}. In the generic case in general relativity there 
is no solution with rotation and without shear. In other words, it has been 
shown that spatial homogeneous, rotating, and expanding universes with 
the perfect fluid have the nonvanishing shear \cite{Ellis73,Ciufolini95}. 
However, there exists such a possibility in the Newtonian
cosmology because some standard obstacles of Newtonian cosmology can 
be overcome in this model \cite{Heckmann61}. The conservation of energy 
gives in a relativistic case
\begin{equation}
\label{eq:3}
\dot{\rho} = - \Theta (\rho + p).
\end{equation}
The occurrence of the term $p$ in the factor ($\rho+p$) has a special
relativistic effect. It is a consequence of the inertia assigned to all
forms of energy. The Newtonian counterpart of equation~(\ref{eq:3}) 
corresponds to $p=0$ (dust for which world lines of fluid are geodesics 
\cite{Ellis73}).

For a rotating, spherically symmetric, self-gravitating system with a 
characteristic size $R$, mean density $\rho$, and rotation $\omega$, 
the angular momentum $J$ is given by $J=2{\cal M} R^2 \omega /5$. Therefore, 
the angular momentum conservation can be written in the Newtonian case 
in the form
\begin{equation}
\label{eq:4}
\omega \rho a^5 = \text{const}
\end{equation}
where $a^3 = a_1 a_2 a_3$ is an average scale factor. Principle (\ref{eq:4})
has its counterpart in the relativistic case \cite{Ellis73}
\begin{equation}
\label{eq:5}
\omega (\rho + p) a^5 = \text{const}.
\end{equation}
In the general relativistic case rotation is governed by a pair of nonlinear
propagation and constraint equations which for the case of the barotropic 
equation of state reduces to \cite{Barrow04}
\[
\dot{\omega_a} = - \left[ \frac{2}{3} \left( 1-\frac{2}{3}c_{s}^{2} \right)
\Theta - \sigma \right] \omega_a
\]
where $\omega_a$ is a shear eigenvector ($\sigma_{ab}\omega^2=\sigma\omega_a$), 
and $c_s$ is a velocity of sound ($c^2_s=\frac{\partial p}{\partial \rho}$).
 
The Einstein equation for the considered case has a first integral in the form
\begin{equation}
\label{eq:6}
\bar{H}^2 = \frac{\Theta^2}{9} = \frac{8\pi G}{3}\rho - \frac{2}{3}\omega^2
+ \frac{\sigma^2}{3} - \frac{{}^3 R}{6} + \frac{\Lambda}{3}
\end{equation}
where ${}^3 R$ is the Ricci scalar of three space equalled to $6k/a^2$ 
($k=-1$ for B(V) and $k=0$ for B(I)) and for a natural system of units 
$8\pi G = c = 1$. We usually work with the equation of state of the form
\begin{equation}
\label{eq:7}
p = (\gamma - 1) \rho
\end{equation}
where $\gamma$ is a constant. From the physical point of view, the most 
important values are $\gamma = 1$ (dust), $\gamma = 4/3$ (radiation), 
$\gamma = 2/3$ (strings), $\gamma_{\lambda} = 2\gamma$ (brane effects), 
$\gamma = 1/3$ (solid dark energy or topological defects). Let us note that 
in equation~(\ref{eq:6}) there is no dependence on a particular form of the 
equation state. 

If the matter is in a pressure-free form it is represented 
only by its 4-velocity $u^a$ and its energy density $\rho$. From the momentum 
conservation we obtain that matter moves along geodesics. For our further 
analysis it would be useful to rewrite
equation~(\ref{eq:4}) to the form of a differential equation
\begin{equation}
\label{eq:8}
\dot{\omega} = - 2 \omega H \qquad \text{or} \qquad
\dot{\omega}=(3\gamma-5)\omega H 
\end{equation}
for the equation of state in the form (\ref{eq:7}). After differentiation 
both sides of equation~(\ref{eq:6}) with respect to the time variable and 
then the substitution of equation~(\ref{eq:6}) we can obtain the Raychaudhuri 
equation. It explicitly demonstrates that in terms of the average Hubble 
function
\[
\bar{H} = \frac{1}{3} (H_1 + H_2 + H_3),
\]
equation (\ref{eq:6}) is really a first integral of basic equation (\ref{eq:1}).
In the considered case we write this equation in the form 
\begin{equation}
\label{eq:9}
\dot{H} = - H^2 - \frac{2}{3} (\sigma^2 - \omega^2) + \frac{\Lambda}{3}
- \frac{1}{6} \rho (3 \gamma - 2),
\end{equation}
where the bar over $H$ will be further omitted. For the FRW symmetry we have 
$H_1=H_2=H_3$, (but in general the Hubble function $H$ is understood to be 
averaged over three different main directions). For anisotropic models B(I) 
and B(V) the average Hubble parameter $\bar{H}$ is calculated from the average
scale factor $\bar{H}=\ln (d/dt (\sqrt[3]{a_1 a_2 a_3}))$. Hence 
equation~(\ref{eq:9}) for the model with the FRW symmetry can be rewritten 
to the form
\begin{equation}
\label{eq:10}
\frac{\ddot{a}}{a} = - \frac{2}{3} (\sigma^2 - \omega^2)
+ \frac{\Lambda}{3} - \frac{1}{6} \rho (3\gamma - 2)
\end{equation}
where
\[
\rho = \rho_0 a^{-3\gamma}, \qquad \sigma = \sigma_0 a^{-3}, \qquad
\omega = \omega_0 a^{-5+3\gamma}.
\]
 
Let us note that from equation (\ref{eq:8}) we obtain that $\omega(a)$ does
not preserve its form in different epochs. The value of $J$ depends on
$\gamma$ ($J = \omega \rho \gamma a^5$). Of course
\[
\frac{\dot{J}}{J} = \frac{\dot{\omega}}{\omega} + \frac{\dot{\rho}}{\rho}
+ 5H = \frac{\dot{\omega}}{\omega} - 3\gamma H + 5H = 0,
\]
and conservation of $J$ takes place if only $\omega$ obeys relation 
(\ref{eq:8}).
 
Therefore the right-hand side of (\ref{eq:10}) can be expressed in terms of
the scale factor $a$ (or average for B(I) and B(V))
\begin{equation}
\label{eq:11}
\ddot{a} = \left( - \frac{2}{3} \sigma_{0}^{2} a^{-6}
- \frac{3\gamma - 2}{6} \rho_0 a^{-3\gamma} + \frac{\Lambda}{3}
+ \frac{2}{3} \omega_{0}^{2} a^{-10+6\gamma} \right) a.
\end{equation}
 
Equation (\ref{eq:11}) can be rewritten in the form analogous to the Newton
equation of motion in the one-dimensional configuration space 
$\{a\colon a \in \mathbf{R}_{+} \}$ where $a$ is the scale factor 
\begin{equation}
\label{eq:12}
\ddot{a} = - \frac{\partial V}{\partial a}
\end{equation}
where the potential function for $\gamma \ne 4/3$ is in the form
\begin{equation}
\label{eq:13}
V(a) = - \frac{\sigma_{0}^{2}}{6} a^{-4} - \frac{\rho_0}{6} a^{-3\gamma+2}
- \frac{\Lambda}{6} a^2 - \frac{1}{3} \frac{\omega_{0}^{2}}{(3\gamma - 4)}
a^{2(3\gamma - 4)} + V_0
\end{equation}
while for $\gamma = 4/3$ it takes the special form
\begin{equation}
\label{eq:14}
V(a) = - \frac{\sigma_{0}^{2}}{6} a^{-4} - \frac{\rho_0}{6} a^{-2}
- \frac{\Lambda}{6} a^2 - C \ln a + V_0
\end{equation}
where $V_0 = \text{const}$ and $C$ is an undetermined integration constant 
in the Hamiltonian which the energy level should be consistent with the 
form of the first integral (\ref{eq:6}).
Equation (\ref{eq:12}) has a first integral which can be expressed for 
$\gamma \ne 4/3$ in terms of the potential function $V(a)$ as
\begin{equation}
\label{eq;15}
V(a) + \frac{\dot{a}^2}{2} = V_0 - \frac{k}{2}.
\end{equation}
Now we can construct the Hamiltonian 
\begin{equation}
\label{eq:16}
\mathcal{H} \equiv \frac{\dot{a}^2}{2} + V(a)
\end{equation}
Trajectories of system (\ref{eq:16}) are situated on the distinguished energy 
level, $\mathcal{H} = E = \text{const}$. We can also define a new Hamiltonian 
flow in the one-dimensional configuration space on the zero energy level by 
translating the curvature contribution to the Hamiltonian. Then
\begin{equation}
\label{eq:17}
\mathcal{H} = \frac{\dot{a}^2}{2} + V(a)= 0
\end{equation}
where for $\gamma \ne 4/3$ 
\begin{equation}
\label{eq:18}
V(a) = - \frac{\sigma_{0}^{2}}{6} a^{-4} - \frac{\rho_0}{6} a^{-3\gamma+2}
- \frac{\Lambda}{6} a^2 - \frac{1}{3} \frac{\omega_{0}^{2}}{(3\gamma - 4)}
a^{2(3\gamma - 4)} + \frac{k}{2}
\end{equation}
and for $\gamma = 4/3$ the Hamiltonian  takes the special form
\[
\mathcal{H}= \frac{\dot{a}^2}{2} + V(a)
= - \frac{\sigma_{0}^{2}}{6} a^{-4} - \frac{\rho_0}{6} a^{-2}
- \frac{\Lambda}{6} a^2 - C \ln a + \frac{k}{2} 
+ \frac{\dot{a}^2}{2} = E = -\frac{1}{3}\omega^{2}_{0}
\]
which becomes consistent with the form of first integral (\ref{eq:6}) only for 
$C=0$. Therefore, for $\gamma \ne 4/3$ the physical trajectories lie on the 
zero-energy level $\mathcal{H} = E = 0$ which coincide with 
the form of the first integral (\ref{eq:18}). For the special case
of $\gamma = 4/3$ trajectories lie on the distinguished energy level
$\mathcal{H} = -\omega^{2}_{0}/3 = \text{const}<0$.
In both cases dynamics is given in the Hamiltonian form and trajectories lie 
on the distinguished level of constant energy.
Finally, for radiation ($\gamma=4/3$) we obtain the new Hamiltonian considered 
on the zero energy level
\begin{equation}
\label{eq:19}
\mathcal{H}= - \frac{\sigma_{0}^{2}}{6} a^{-4} - \frac{\rho_0}{6} a^{-2} 
+ \frac{\dot{a}^2}{2} - \frac{\Lambda}{6} a^2 + \left( \frac{k}{2} 
+ \frac{1}{3}\omega_{0}^{2} \right) \equiv 0.
\end{equation}

Note that this model is dynamically equivalent to standard model 
with some effective curvature, 
\begin{equation}
\label{eq:19b}
k_{\text{eff}} = k + \frac{2}{3} \omega_{0}^{2}.
\end{equation}
Let us consider the FRW ($\sigma_0 = 0$)
model filled with some pure unknown matter $X$ such that $p_X = w_X \rho_X$.
Then we obtain
\begin{equation}
\label{eq:19c}
V(a) = - \frac{\rho_0}{6} a^{-3w_X +2} + \frac{\Lambda}{6} a^2 -
\frac{\omega_0^2}{3(3w_X-4)} a^{2(3w_X-4)} + \frac{k}{2}.
\end{equation}
The mixture of noninteracting dust and unknown matter $X$ satisfies the
equation of state
\begin{equation}
\label{eq:20}
p=p_X+0=(\gamma(a)-1)\rho=
\frac{w_X}{\frac{\rho_{m0}}{\rho_{X0}}a^{3(w_X -1)}+1}\rho, \quad
\rho=\rho_{m0}a^{-3}+\rho_{X0}a^{-3w_X}.
\end{equation}
Then the potential function is given by the integral
\begin{equation}
\label{eq:21}
V(a) = \int\limits^a \left[ \frac{3(\gamma(a) - 2)}{6}
+ \frac{2}{3}\sigma^{2}(a) - \frac{\Lambda}{3}
- \frac{2}{3}\omega_0^2 a^{-10}
\exp\left(6\int\limits^a \frac{\gamma\left(a\right)}{a} da\right) \right] a\,da
\end{equation}
where $\gamma(a)$ is derived from equation~(\ref{eq:20}) and for the mixture 
of dust and unknown matter we can find that
\begin{equation}
\label{eq:21b}
\int\limits^a \frac{\gamma(a)}{a} da = \int\limits^a
\frac{w_X \epsilon_{0x} a^{m-1}}{\epsilon_{0X}a^m +\epsilon_{0m}a^{-3}}da=
\frac{w_X}{m+3} \ln(\epsilon_{0m}+\epsilon_{0X}a^{m+3})
\end{equation}
where $m = -3(1 + w_X)$. Finally we obtain
\begin{equation}
\label{eq:21c}
V(a) = - \frac{\rho_{X0}}{6} a^{-3w_X + 2} - \frac{\sigma_0^2}{6} a^{-4}
- \frac{\Lambda}{6} a^2 + \frac{k}{2} - \frac{2}{3}\omega_0^2
\int\limits^a \frac{a^{-9} da}{(\epsilon_{0m}+\epsilon_{0X}a^{m+3})^2}
\end{equation}
 
Let us consider for example the case of solid dark energy
($w_X = 1/3$, $m = -4$). To do this we substitute these values into
integral (\ref{eq:21b}) which gives
\[
\tilde{J} =
\int\limits^a \frac{da}{a^8 (\epsilon_{0m}+\epsilon_{0X})^2} =
- \frac{1}{\epsilon_{0m} \epsilon_{0X}} \left[ \frac{1}{a^8(1 + \alpha a)}
+ 8\sum\limits_{k=1}^6 \frac{(-1)^k \alpha^{k-1}}{(7-k)!a^{7-k}}
- \alpha^6 \ln \frac{1+\alpha a}{a} \right]
\]
where $\alpha={\epsilon_{0m}/\epsilon_{0X}}$.
 
Then we obtain potential in term of the integral $\tilde{J}$
\begin{equation}
\label{eq:21d}
V(a) = - \frac{\rho_{X0}}{6}a - \frac{\sigma_0^2}{6}a^{-4}
- \frac{\Lambda}{6}a^2 + \frac{k}{2} - \frac{2}{3}\omega_0^2 \tilde{J}.
\end{equation}

\section{Exact solution with rotation}

In the case of the dust matter, an exact solution with rotation seems to be 
especially interesting. It can be easily shown that it can be simply obtained
if we note that the corresponding dynamical equation reduces exactly to the 
best known case of the FRW model with dust, radiation, and the cosmological
constant. 

In the case of vanishing shear and the cosmological constant $\Lambda=0$ 
an exact solution can be given in the simplest form. There are three 
possibilities of positive ($k=+1$), zero ($k=0$), and negative ($k=-1$) 
curvature.

a) If $k=+1$ equation~(\ref{eq:17}) becomes
\[
\dot{a}^2 = -2V(a) = \frac{\rho_0}{3} \frac{1}{a^{3\gamma-2}} +
\frac{2}{3} \frac{\omega_0^2}{(3\gamma-4)a^{-6\gamma+8}} - 1,
\]
and since $\dot{a}^2$ has to be positive there are maximum and minimum values
for $a$ (if $\gamma<4/3$) as can be derived using elementary techniques. The
general solution for $\gamma=1$ (dust) is
\[
t = -(\beta+\alpha a-a^2)^{1/2} - \frac{\alpha}{2}
\arcsin{\frac{\alpha-2a(t)}{(\alpha^2 +4\beta)^{1/2}}} + C,
\]
where $\alpha=\rho_0/3$, $\beta=-\frac{2}{3}\omega_0^2$ and
$a_{\text{min}}<a<a_{\text{max}}=\frac{1}{2}((\alpha^2+4\beta)^{1/2}+\alpha)$, 
$\omega_0 \le \rho_0 \sqrt{2}$.
 
Since only the first derivative vanishes for $a=a_{\text{max}}$, we obtain
an oscillating singularity-free universe.

b) If $k=0$ equation (\ref{eq:17}) gives
\[
\dot{a}^2=\frac{\alpha}{a^{3\gamma-2}}+\frac{\beta}{a^{-6\gamma+8}}.
\]
Here $\dot{a}^2$ is always positive for $a > - \beta/\alpha$.
So there is no maximum value for $a$. The result is a monotonically
expanding universe without an initial singularity.
 
For $\gamma$=1 the general solution is
\[
t=\frac{2}{3\alpha^2}(\alpha a + \beta)^{1/2}(\alpha a-2\beta)+C.
\]
 
c) The last case, $k=-1$, gives the equation
\[
\dot{a}^2=\frac{\alpha}{a^{3\gamma-2}}+\frac{\beta}{a^{-6\gamma+8}}+1.
\]
This results is also a monotonically expanding universe without an initial
singularity.
 
The scale factor has a minimum 
\[
a \ge a_{\text{min}}=\frac{-\alpha+\sqrt{\alpha^2-4\beta}}{2}.
\]
 
For $\gamma=1$ the solution of (\ref{eq:17}) is
\[
t=(a^2+\alpha a+\beta)^{1/2}-\frac{\alpha}{2}\ln{(2(a^2+\alpha a+\beta)+
2a+\alpha)}+C,
\]
if $\alpha^2-4\beta>0$,
\[
t=(a^2+\alpha a+\beta)^{1/2}-\frac{\alpha}{2}\ln{(2a+\alpha)}+C,
\]
if $\alpha^2-4\beta=0$,
\[
t=(a^2+\alpha a+\beta)^{1/2}-\frac{\alpha}{2}\ln{(2a+\alpha)}+C.
\]
 
Exact formulas for observables are also possible. For example, in the case of 
$\Lambda=0$ the luminosity distance $D_0$ between a galaxy, which emitted 
light at the cosmic moment $t$ and the observer who detected the light at 
the present moment $t_0$, is given by
\[
D_0 = \frac{\Omega_{\text{m},0}z + \left(\Omega_{\text{m},0}
+ 2\Omega_{\omega,0}-1\right)\{[2\Omega_{\omega,0}z^2
+ 2z\left(2\Omega_{\omega,0} + \Omega_{\text{m},0}\right)+1]^{1/2}-1\}}%
{H_0[\Omega_{\text{m},0}^2 + 2\Omega_{\omega,0}\left(2\Omega_{\omega,0}
+ 2\Omega_{\text{m},0}-1\right)]}.
\]
The above formula is valid for all values of curvature index
(see for details \cite{Dabrowski86}).
 
In the more general case of the nonvanishing cosmological constant we can
find also a solution in the term of an elliptic function. In the case of dust 
matter, the problem of finding an exact solution is equivalent to integration 
of the equation
\[
\left(\frac{dx}{dt}\right)^2 = \Omega_{k,0} + \Omega_{\text{m},0} x^{-1}
+ \Omega_{\omega,0} x^{-2} + \Omega_{\Lambda,0} x^2,
\]
which takes the following form in the conformal time 
\[
\left(\frac{dx}{d\eta}\right)^2 = \Omega_{\omega,0} + \Omega_{\text{m},0} x
+ \Omega_{k,0} x^2 + \Omega_{\Lambda,0} x^4 \equiv W(x),
\]
where $d\eta \equiv dt/x$, $x \equiv a/a_0$ and $a_0$ is a present value of
the scalar factor $a$.
 
Then it can be integrated explicitly using the Weierstrass $\wp$ function,
to yield 
\[
\wp(\eta) = \frac{\sqrt{W(x)W(x_0)}+W(x_0)}{2(x-x_0)^2} +
\frac{4 l x_0^3 + 2 k x_0 + m}{4(x-x_0)} + \frac{6 l x_0^2 + k}{12}
\]
or equivalently
\[
x-x_0 = \frac{\sqrt{W(x_0)}\wp'(\eta) +
\frac12[\wp(\eta)-\frac{6\Omega_{\Lambda,0}x_0^2
+\Omega_{k,0}}{12}](4\Omega_{\Lambda,0}x_0^3+2\Omega_{k,0}x_0+\Omega_{m,0})+
W(x_0)\Omega_{\Lambda,0}x_0}{2[\wp(\eta)-\frac{6\Omega_{\Lambda,0}x_0^2+
\Omega_{k,0}}{12}]^2-\frac12\Omega_{\Lambda,0}W(x_0)}
\]
where $x_0 = x(\eta=0)$, and $\wp$ is constructed with the invariants
$g_2 = \Omega_{\Lambda,0} \Omega_{\omega,0} + \frac{1}{12}\Omega_{k,0}^2$,
$g_3 = \frac16 \Omega_{\Lambda,0}\Omega_{k,0}\Omega_{\omega,0} -
(\Omega_{k,0}/6)^3 - \frac{1}{16}\Omega_{\Lambda,0}\Omega_{\text{m},0}$.

In the considered case the effect of the global rotation is modelled by
negative radiation. Then, if we put in the place of dimensionless 
$\Omega_{r,0}$ some effective constant $\Omega_{\omega, 0}^{\text{eff}}$,
(which combine all contribution coming from matter scaling like radiation)
we immediately obtain the exact solution in terms of the Weierstrass elliptic 
function \cite{Dabrowski86}. It is interesting that in this case formulas 
determining the magnitude-redshift relation can be also given in the exact 
form. After introducing $k_{\text{eff}}$ (see formula (\ref{eq:19b})) we can 
also simply obtain the corresponding exact formulas from that obtained for 
$\Omega_{k,0}\ne 0$.

\section{Cosmic acceleration and horizon in the Universe with global rotation}

In our further analysis we will explore dynamics given by the canonical
equations. Assuming $x = a$ and $y = \dot{a} = \dot{x}$ we have from
equation (\ref{eq:17})
\begin{subequations}
\label{eq:22}
\begin{align}
\label{eq:22a}
\dot{x} &= \frac{\partial \mathcal{H}}{\partial y} = y \\
\label{eq:22b}
\dot{y} &= - \frac{\partial \mathcal{H}}{\partial x}
= - \frac{\partial V}{\partial x}.
\end{align}
\end{subequations}
with the first integral in the form $y^{2}/2+V(x) \equiv 0$. Then, we can 
perform a qualitative analysis of autonomous system (\ref{eq:22})
in the phase plane $(a, \dot{a}) = (x, y)$.

The general observation is that if we consider the two-dimensional Hamiltonian
dynamical system then eigenvalues of the linearization matrix
$\lambda_1,\lambda_2 \colon
\lambda^2 = - \frac{\partial^2 V}{\partial a^2}\vert_{a_0}$.
They are saddle points if $\frac{\partial^2 V}{\partial a^2}\vert_{a_0} <0$ 
and then eigenvalues are real of opposite signs otherwise they are centres 
at which eigenvalues are purely imaginary and conjugated.
This dynamical system rewritten to more useful form will be considered in 
section 5.

We can also  observe that trajectories of system (\ref{eq:22}) can be 
integrable in quadratures, namely,
from the Hamiltonian constraint $\mathcal{H} = E = 0$, we obtain
\begin{equation}
\label{eq:23}
t-t_0 = \int\limits_a \frac{da}{\sqrt{-2V(a)}}.
\end{equation}

Note that it is possible to make the classification of
qualitative evolution paths by analyzing the characteristic curve which
represents the boundary of domain admissible for motion. For this 
purpose we consider the equation of zero velocity, $\dot{a} = 0$ which 
represents the boundary in the configuration space. Because
\begin{equation}
\label{eq:24}
\dot{a}^2=2V(a)
\end{equation}
the motion of the system is limited to the region
$\{a \colon V(a) \leq 0 \}$. Let us consider the boundary set of the
admissible for motion configuration space given by a condition
\begin{equation}
\label{eq:25}
\partial M \equiv \{ a \in \mathbf{R}_{+} \colon V(a)=0 \}.
\end{equation}
For the special case of $\gamma =4/3$ of course levels $V(a) = E$
should be considered. From the exact form of the potential
function (\ref{eq:18}) or (\ref{eq:21}) parameters $\Lambda$
or $\omega_0$ can be expressed as a function of $a$.
For example from (\ref{eq:18}) for $\gamma \ne 4/3$ we obtain
\begin{equation}
\label{eq:26a}
\Lambda(a) = \frac{6}{a^2}
\left[ - \frac{\sigma_{0}^{2}}{6} a^{-4} - \frac{\rho_0}{6} a^{-3\gamma+2}
- \frac{1}{3} \frac{\omega_{0}^{2}}{(3\gamma - 4)}
a^{2(3\gamma - 4)} + \frac{k}{2} \right]
\end{equation}
and for $\gamma = 4/3$
\begin{equation}
\label{eq:26b}
\Lambda(a) = \frac{6}{a^2}
\left[ - \frac{\sigma_{0}^{2}}{6} a^{-4} - \frac{\rho_0}{6} a^{-2}
- \frac{2}{3}\omega_0^2 + \frac{k}{2} \right].
\end{equation}
Finally, we consider the evolution path as a level of $\Lambda = \text{const}$
and then we classify all evolution scenarios modulo with their quantitative 
properties of dynamics (to compare see \cite{Robertson33,Dabrowski96}). For 
the special case of radiation the corresponding classification of the FRW 
model effective curvature $k_{\text{eff}}$ can be achieved in a similar way.
Then we obtain the analogous phase portraits is shown on Fig.~\ref{fig:1}.
 
From the physical point of view it is interesting to answer the questions:
are the trajectories distributed in the phase space in such a way that
critical points are typical or exceptional? How are trajectories with
interesting properties distributed? For example, along which trajectories
the acceleration condition, $\ddot{a} =-dV/da > 0$ is satisfied?
In order to fulfil that $V(a)$ must be decreasing. One can easily observe this
phenomenon from the geometry of the potential function. In the phase space,
the area of accelerations is determined by condition $\dot{y} >0$ or by 
corresponding condition in the configuration space
\begin{equation}
\label{eq:27}
\frac{2}{3} \sigma_{0}^{2} a^{-6} + \frac{3\gamma - 2}{6} \rho_0 a^{-3\gamma}
- \frac{\Lambda}{3} - \frac{2}{3}\omega_{0}^{2} a^{-10+6\gamma} < 0.
\end{equation}
The presence of the last term in inequality (\ref{eq:27}) demonstrates 
that in the configuration space the domain of acceleration is larger than for 
the case of the FRW model with vanishing global rotation. Let us note that 
the positive value of the cosmological constant $\Lambda$ acts in the same 
direction as rotation, i.e., global rotation acts as dark energy (or dark 
radiation).

It can easily be demonstrated for the FRW models that if
$\dot{a}(t) \to \text{const}$ as $a \to 0$ then the
model has no particle horizons (in the past) \cite{Weinberg72}.
Indeed, if there exists a constant $C$ such as for sufficiently
large $\rho$, the velocity of the scale factor is upper bounded 
$da/dt \le C$ and
\begin{equation}
\label{eq:28}
\int\limits_0^{a_0} \frac{da}{a} < C\int\limits_0^{t_0} \frac{dt}{a}.
\end{equation}
The integral on the left hand-side of (\ref{eq:28}) diverges and there are 
no causally disconnected regions. Putting this in terms of $(x,y)$ variables 
one needs the condition, as $x \to 0$, $y \to \text{const}$, as the sufficient 
condition for solving the 
horizon problem. Now, from the Hamiltonian constraint we obtain that as 
$x \to 0$, then $V(x) \to \text{const}$ (zero is included).

\section{The phase plane analysis---general properties of the models}

Let us apply a dynamical system method to analyse the system under
consideration.
The dynamics of cosmological models with global rotation, the cosmological
constant $\Lambda$ and matter, satisfying the equation of state (\ref{eq:7}),  
is described by the following dynamical system
\begin{subequations}
\label{eq:29}
\begin{align}
\label{eq:29a}
\dot{a} &= y \\
\label{eq:29b}
\dot{y} &= - \frac{2}{3} \sigma_{0}^{2} a^{-5}
- \frac{3\gamma-2}{6} \rho_0 a^{-3\gamma+1}
+ \frac{\Lambda}{3} a + \frac{2}{3} \omega_{0}^{2} a^{6\gamma-9}.
\end{align}
\end{subequations}

Let us note that some general properties of system (\ref{eq:29}) for 
vanishing shear ($\sigma_0 = 0$) and the matter in the form of solid dark 
energy (or topological defects) with $\gamma=1/3$. Then when $a \to 0$ 
the effect of the cosmological constant is negligible.
In this case $\dot{y} = \frac{1}{6} \rho_0 + \frac{2}{3}\omega_{0}^{2} a^{-7}$
and the term with rotation cannot dominate the matter term with 
positive energy because $V(a) \le 0$. From the existence of the first 
integral we have
\[
y=\sqrt{-2V(a)} = \sqrt{\frac{1}{3}\rho_0 a + \frac{2}{3}\omega_{0}^{2}
a^{-5} - k} \qquad \text{and} \qquad
a>\left( \frac{2}{3\rho_0}\omega^2_0\right)^{1/7}.
\]
It is possible that density of solid dark energy and the rotation compensate
each other, when the scale factor reaches $a=a_{\text{min}}$ near the
singularity.

As $a \to 0$ we have $\dot{y}=\frac{1}{6}\rho_0$, i.e., $y\sim t$ and 
$a\sim t^2$ and we obtain that the horizon problem is solved because 
$y=\dot{a} \to 0$ as $t \to 0$. The phase portrait for this case show 
Fig.~\ref{fig:3}.
 
To apply the method of dynamical system let us consider the FRW model 
with dust matter, the cosmological constant, and global rotation. 
Then the corresponding dynamical system has the form
\begin{subequations}
\label{eq:30}
\begin{align}
\label{eq:30a}
\dot{a} &= y \\
\label{eq:30b}
\dot{y} &= - \frac{1}{6}\rho_0 a^{-2} + \frac{\Lambda}{3} a
+ \frac{2}{3} \omega_{0}^{2} a^{-3}
\end{align}
\end{subequations}
with the first integral, for the general form of the equation of state  
$p = (\gamma - 1) \rho$, given by 
\begin{equation}
\label{eq:31}
y^2 = \frac{1}{3}\rho_0 a^{-1} + \frac{\Lambda}{3} a^2 - k
- \frac{2}{3} \omega_{0}^{2} \frac{a^{2(3\gamma-4)}}{3\gamma-4},
\end{equation}
 
Near the initial singularity of type, i.e., as $a \to \text{const}$ 
the generic flat solution without the cosmological constant has the form
\begin{equation}
\label{eq:31a}
t-t_0 = \frac{2}{3\rho^{2}_{0}} \left( \frac{\rho_0}{3} a
+ \frac{4}{3} \omega^{2}_{0} \right)
\sqrt{\frac{\rho_0}{3} a + \frac{2}{3}\omega^{2}_{0}}.
\end{equation}
 
If we assume that $\omega_0 \ll 1$ then from solution (\ref{eq:31a}) 
we obtain a simple monotonic solution with the singularity at 
$a=a_0=\frac{2\omega^2_0}{\rho_0}$ 
\begin{equation}
\label{eq:31b}
t-t_0 = \frac{2a}{9\sqrt{3\rho^{2}_{0}}}
\sqrt{a + \frac{2\omega^{2}_{0}}{\rho_0}}.
\end{equation}

Note, that as in the previous case there is the possibility that the scale 
factor reaches $a=a_{\text{min}}$ in the initial singularity. From the 
condition that motion is admissible in the region $V(x) \le 0$ we have for 
the flat model the condition
\[
\forall x \colon -\frac{1}{2} \Omega_{\text{m},0}x^{-1}
-\frac{1}{2} \Omega_{\omega,0}x^{-2}<0
\]
or in terms of redshift $z$
\[
\forall z \colon  \Omega_{\text{m},0}+ \Omega_{\omega,0}(1+z)>0.
\]

For the minimum $a=a_{\text{min}}$ and the maximum $\rho_\text{m}$ 
we finally obtain the bound
\[
-\Omega_{\omega,0} < \frac{\Omega_{\text{m},0}}{1+z_{\text{sing}}},
\]
where $\Omega_{\text{m},0} \simeq 0.3$ and $z_{\text{sing}}$ is the value 
of redshift at the moment of singularity $a=a_{\text{min}}$ such that 
$1+z_{\text{sing}}=\frac{a_0}{a_{\text{min}}}$.
It would be useful to compare the present effects of radiation and rotation 
because both scale in the same way. The rotation effect could dynamically 
dominate radiation i.e.,
\[
-\Omega_{\omega,0}>\Omega_{r,0} \simeq 0.5 \cdot 10^{-4}.
\]
 
The contribution coming from both radiation and rotation we denote as
$\Omega^{\text{total}}_{r,0}$. Then, from equation~(\ref{eq:30b}) in the case 
without the cosmological constant it can be obtained the relation
\[
\ddot{x}=-\frac{\Omega_{\text{m},0}}{2} x^{-2} 
\left(1+\frac{2\Omega^{\text{total}}_{r,0}}{\Omega_{\text{m},0}}x^{-1}\right).
\]
Therefore, for $x_0=-\frac{2\Omega^{\text{total}}_{r,0}}{\Omega_{\text{m},0}}$
we have an inflection point on diagram $x(t)=\frac{a}{a_0}$ which separates 
domains in which $\ddot{x}$ has different convexity. 

Our starting point for further analysis of results will be dynamical equations 
rewritten to the new form using dimensionless quantities:
\begin{gather}
\label{eq:32}
x \equiv \frac{a}{a_0}, \qquad T \equiv \vert H_0 \vert t, \qquad
\Omega_{i,0} \equiv \frac{\rho_{i,0}}{\rho_{\text{cr},0}}, \\
\rho = \rho_0 \left( \frac{a}{a_0} \right)^{-3\gamma}, \qquad
\sigma = \sigma_0 \left( \frac{a}{a_0} \right)^{-3}, \qquad
\omega = \omega_0 \left( \frac{a}{a_0} \right)^{-5+3\gamma},
\end{gather}
with $H=\frac{\dot{a}}{a}$, $\rho_{\text{cr}}= 3{H_0}^2$ where the index $0$ 
denotes present day values (at time $t_0$). After introducing the density 
parameter of rotation for $\gamma \neq 4/3$ the dynamical effect of global 
rotation is equivalent to some kind of additional, noninteracting fluid
for which we have
\begin{gather}
\label{eq:33}
\rho_{\omega} = \frac{2\omega_{0}^{2}}{3\gamma -4} x^{2(3\gamma - 5)}
= \rho_{\omega_0} x^{6\gamma-10} , \qquad
\Omega_{\omega} = \frac{\rho_{\omega}}{\rho_{cr}}
= \Omega_{\omega,0} x^{6\gamma-10}, \\
w_{\omega} = \frac{10-6\gamma}{3}, \qquad
p_{\omega} = (w_{\omega} - 1) \rho_{\omega},
\end{gather}
and negative energy and in consequence the negative density parameter 
$\Omega_{\omega}$. Hence, the basic dynamical equations are 
\begin{equation}
\label{eq:33b}
\frac{\dot{x}^2}{2} = \frac{1}{2}\Omega_{k,0}
+ \frac{1}{2} \sum\limits_{i=1} \Omega_{i,0} x^{2-3w_i} \\
\end{equation}
\begin{subequations}
\label{eq:34}
\begin{align}
\label{eq:34a}
\dot{x} &= x \\
\label{eq:34b}
\dot{y} &= \ddot{x} = \frac{1}{2} 
\sum\limits_{i=1} \Omega_{i,0} (2-3w_i) x^{1-3w_i} 
= - \frac{\partial V}{\partial x}.
\end{align}
\end{subequations}
where for $w_i \ne 2/3$ ($\gamma \ne 4/3$)
\begin{gather}
V(x) = - \frac{1}{2} \Omega_{k,0} - \frac{1}{2}
\sum\limits_{i=1} {\Omega_{i,0} x^{2-3w_i}}, \\
\frac{1}{2} \Omega_{\omega,0} = \frac{2}{3}
\frac{\omega_{0}^{2}}{(3\gamma - 4)}, \nonumber
\end{gather}
and the system is now defined on the zero energy level $\mathcal{H} = E = 0$.
 
If  $w_i = 2/3$ ($\gamma = 4/3$) we have
\begin{equation}
V(x) = \frac{1}{2} \sum\limits_{w_i \ne 1} {\Omega_{i,0} x^{2-3w_i}}
+ \frac{1}{3} \omega^2_0,
\end{equation}
where in general $p_X = (w_X - 1) \rho_X$ and for unknown dark energy $X$, 
$w_1 = 1/3$ (solid dark energy), $w_2 = 2/3$ (effects of string), 
$w_3 = \gamma$ (perfect fluid), $w_{\Lambda} = 0$ (cosmological constant),
$w_{\omega} = \frac{10-6\gamma}{3}$ (for rotation), 
$w_{\lambda} = 2\gamma$ (effects of brane).

We could in principle include the curvature term $\frac{1}{2}\Omega_{k,0}$ 
into the sum in the right hand sides of (\ref{eq:33b}),
it would correspond to $w_k = 2/3$.
We see that in particular the case $x_0=1$, $\dot{x}_0 = 1$ is a solution.
From this definition $\ddot{x}_0 = - q_0$, where $q_0$ is the deceleration
parameter. Therefore, $\sum\limits \Omega_{i,0}+ \Omega_{k,0}=1$
(presently but this equality can be extended to any given time).
For the special case of the universe filled with radiation matter we can also 
formally define energy density of some fictitious fluid which mimics the 
rotation effect and scales like ``curvature fluid'', 
$\rho_{\omega}=-2\omega^2_0 a^{-2}$.
Hence, the corresponding density parameter is
\[
\Omega_{\omega} = \frac{\rho_{\omega}}{3H_{0}^{2}}, \qquad
\]
and because the system is defined on the levels $\mathcal{H}=E=0$ we have 
the constraint $\sum_{i} \Omega_{i,0} = 1$.

Therefore in any case the system under consideration can be represented in the 
form of the multifluid FRW dynamics.
Of course, dynamical system  (\ref{eq:34}) is the Hamiltonian
\begin{equation}
\label{eq:35}
\mathcal{H} = \frac{p_{x}^{2}}{2} + V(x)
\end{equation}
where $p_x$ is the momentum conjugated with the generalised coordinate $x$. 

In the following we consider a universe filled with dust and some unknown
component $X$ with negative pressure called ``dark energy'', so that the 
strong energy condition is violated $\rho_X + 3p_X <0$. We assume that the 
weak energy condition $\rho_X+p_X>0$ is satisfied, so $w_X >0$ for the
hypothetical quintessence matter.
Hence, the Universe is accelerating provided that the following condition is
satisfied
\begin{equation}
\label{eq:36}
2\ddot{a}(x) = \Omega_{X,0}(2-3w_X) x^{1-3w_X} - \Omega_{\text{m},0} x^{-2}
+ 2\Omega_{\omega,0}(3\delta(x) -4) x^{3(2\gamma-3)}
+ 2\Omega_{\Lambda,0} x > 0,
\end{equation}
where
\[
\delta(x) = \frac{w_X}{\frac{\Omega_{\text{m},0}}{\Omega_{X,0}}x^{3(w_X-1)}
+ 1}, \qquad p=\delta(x)\rho
\]
and $p,\rho$ are total pressure and energy density; $p=0+w_X\rho_X$,
$\rho=\rho_m+\rho_X$.

Our universe accelerates at present ($x=1$) if
\begin{equation}
\label{eq:37}
\Omega_{X,0}(2-3w_X) - \Omega_{\text{m},0}
+ 2\Omega_{\omega,0}(3\delta(1) - 4) + 2 \Omega_{\Lambda,0} > 0.
\end{equation}
Let put $\Omega_{\Lambda,0} = 0$ and matter $X$ denotes a candidate for the 
dark energy (the pure cosmological constant corresponds to $w_X = - 1$).

Let $\sigma =\Omega_{\text{m},0}/\Omega_{X,0}$,
$\delta = \Omega_{\omega,0}/\Omega_{X,0}$, then is required
\begin{equation}
\label{eq:38}
w_X < \frac{(8\delta+\sigma - 2)(\sigma+1)}{3(2\delta - \sigma - 1)}.
\end{equation}
If $\sigma = 3/7$ then for small $\delta$ we have
\begin{equation}
\label{eq:39}
w_X \le \frac{11}{21}(1 + 40.6 \vert \delta \vert), \qquad
\delta < 0.
\end{equation}
 
Therefore, the corresponding conditions of negativeness of coefficient
of state for dark energy is weaker then for the case of vanishing
rotation. Present experimental estimates are based on estimation of barions in
clusters of galaxies giving $\Omega_{\text{m},0} \sim 0.3$ \cite{Peebles03}.
On the other hand, the location of the first acoustic peak in the CMB detected 
by the experiments Boomerang, Maxima and WMAP suggest a nearly flat
universe. It also implies that our Universe is
presently accelerating for a wide range of values of coefficients of the 
equation of state, roughly $w_X < 0.5$. Let us note that now we have the 
constraint on the angular velocity
\begin{equation}
\label{eq:40}
\omega_0 = \sqrt{\frac{3\left(
\Omega_{k,0}+\Omega_{\text{m},0}+\Omega_{\Lambda,0}-1\right)}{2}} H_{0}
= \sqrt{\frac{-3\Omega_{\omega,0}}{2}} H_{0}
\end{equation}
which is obtained as a consequence of the constraint relation 
$\Sigma \Omega_{i,0}=1$.
 
It is easy to obtain that for $\Omega_{\omega,0} \sim 0.01$ which corresponds
to the limit obtained by us from SNIa data
and the present value $\rho_0 = 0.8 \cdot 10^{-29}$ g cm$^{-3}$,
$\omega_0 \sim 2.6 \cdot 10^{-19}$ rad s$^{-1}$ 
that is in a good agreement with the observational limit \cite{Godlowski04}.

Let us now make some important remarks concerning acceleration regions 
transition in the configuration space admissible for motion 
$x_{\text{min}} < x \le \infty$. The ``physical region'' is 
$x_{\text{min}} < x \le 1$ because $x \equiv a/a_{0}$. Accelerated 
expansion starts from same $x_{a}$ given by non-zero solutions of the equation
\[
h(x) = - \frac{\partial V(x)}{\partial x}x^3 = - \Omega_{\text{m},0} x
+ 2 \Omega_{\Lambda,0} x^4 - 2 \Omega_{\omega,0} = 0
\]
which represents the boundary of acceleration region {$h(x) \ge 0$} in the
configuration space.
 
The function $h(x)$ satisfies the condition $h''(x)>0$ if only 
$\Omega_{\Lambda,0}>0$ and for large and small $x$ where 
it has asymptotic forms $h(x) \propto \Omega_{\Lambda,0} x^4$
and $h(x) = - \Omega_{\text{m},0} x - 2 \Omega_{\omega,0}$, respectively.
Additionally, we have $h(0) = - 2 \Omega_{\omega,0} >0$. Therefore,
$h(x)$ has a minimum at 
$x_{\text{min}} = \sqrt[3]{\Omega_{\text{m},0} / 2 \Omega_{\Lambda,0}}$
which coincides with the same value for the standard case  with
vanishing $\Omega_{\omega,0}$.

Let us consider the model without rotation $\Omega_{\omega,0} = 0$. 
Then $x \in [0,1]$. Now, we can simply define the fraction of acceleration 
during the whole evolution up today as
\[
P = \frac{1 - x_{a}}{1} = 1 - \sqrt[3]{\frac{\Omega_{\text{m},0}}
{2 \Omega_{\Lambda,0}}},
\]
where $x_a=\frac{a_{\text{tr}}}{a_0}$ is the value of the relative scale 
factor in the transition epoch $a_{\text{tr}}$.
For these models accelerated expansion starts at $x_{a}$,
which corresponds to transition redshift
\[
z_{a} = x_{a}^{-1} - 1 = \sqrt[3]{\frac{2 \Omega_{\Lambda,0}}
{\Omega_{\text{m},0}}} - 1.
\]
The fact that $z_{a}$ is so close to zero (for the FRW model with 
the cosmological constant 
term, $z_{a} = 0.414$) is the cosmic coincidence 
problem---$z_{a}$ is shifted towards smaller redshift.
 
Let us note that this obstacle is less dramatic in the class of models with 
rotation $\Omega_{\omega,0} \ne 0$, because due to presence of rotation
contribution $z_{a}$ is shifted towards larger redshift.

\section{Discussion and detailed analysis of the phase plane}

Dynamical system methods of analysis dynamics of the FRW model with global
rotation seems to be useful if we are interested in the qualitative
properties of dynamics and influence of global rotation on the dynamics of 
the universe.
In this approach we deal with the full global dynamics of the universe
whose asymptotic states are represented by critical points of the
systems. In such a representation the phase diagrams in a two-dimensional
phase space allow us to analyse the acceleration in a clear and natural way.
It is the consequence of representation of dynamics as a one-dimensional
Hamiltonian flow.
 
From the theoretical point view it is important to know how large the class 
of accelerated models is, or what kind of qualitative behaviour of 
trajectories introduce the global rotation. We will call this class of 
accelerated models typical (or generic), if the domain
of acceleration in the phase space driven by effect of global rotation is
non-zero measure. On the other hand, if only nongeneric (or zero measure)
trajectories are represented by accelerated universes, then the mechanism
which drives these trajectories to accelerate should be called ineffective. 
Such a point of
view is a consequence of the fact that, if the acceleration is an attribute
of a trajectory which starts with a given initial condition it should also be 
an attribute of trajectories which start from nearby initial conditions.
From recent astronomical measurements we obtain 
that while global rotation effects give rise to the acceleration of the 
universe, the cosmological constant is still needed to explain its rate.

The system under consideration is
\begin{subequations}
\label{eq:51}
\begin{align}
\label{eq:51a}
\dot{x} &= y \\
\label{eq:51b}
\dot{y} &= - \frac{1}{2} \sum_{i} \Omega_{i,0} (3w_i - 2) x^{1-3w_i}
= - \frac{\partial V}{\partial x}
\end{align}
\end{subequations}
with first integral
\[
\frac{y^2}{2}  - \frac{1}{2} \sum_{i} \Omega_{i,0} x^{2-3w_i}=0.
\]
In general, it represents the dynamics of FRW models filled with
noninteracting multifluid for each $i$ component in the equation of state
$p_i = (w_i - 1)\rho_i$ is satisfied.
 
\begin{figure}
\includegraphics[width=0.7\textwidth,angle=-90]{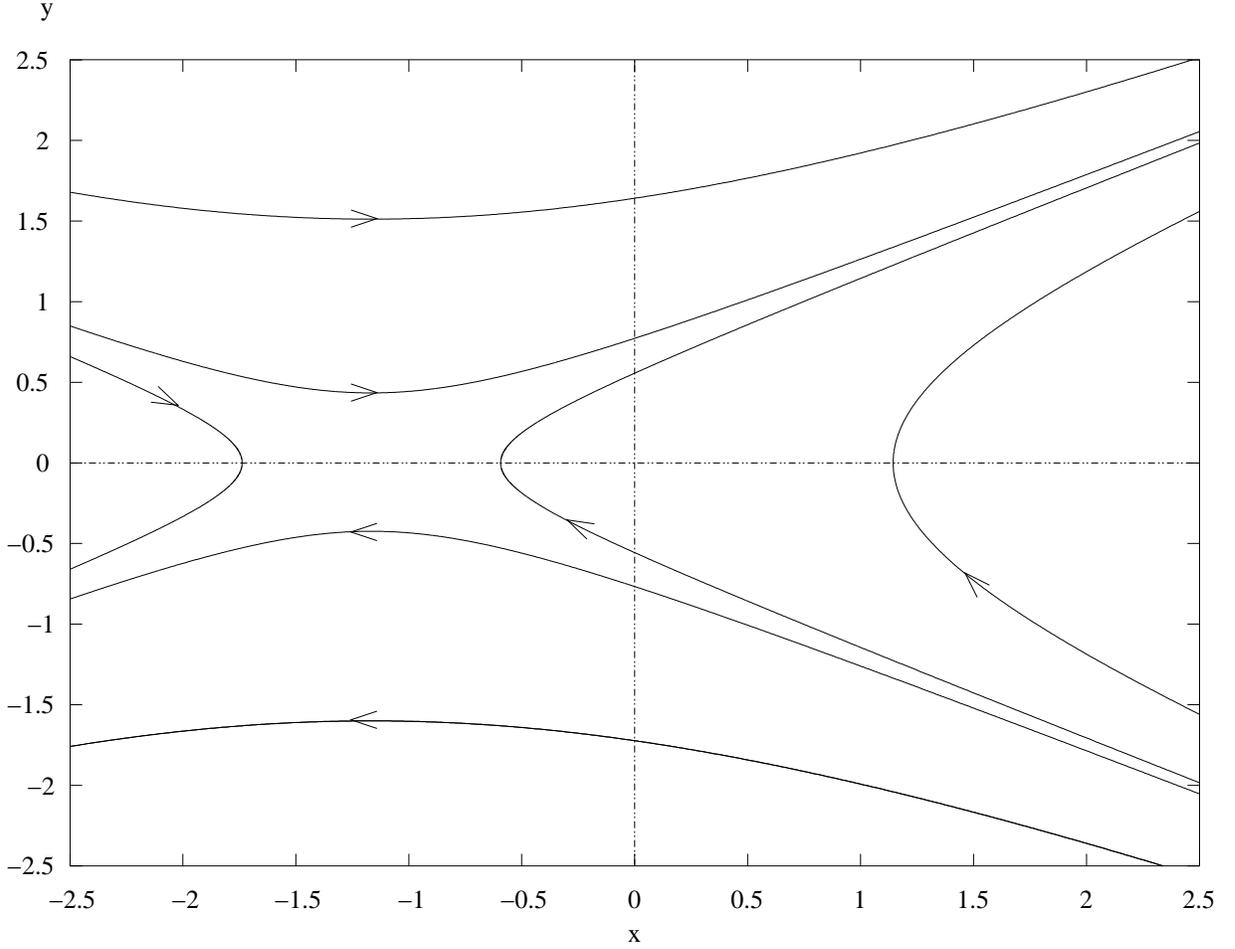}
\caption{
The phase portrait of system for models with global rotation $\Omega_{\omega,0} \ne 0$
and dark matter $X$ in the form of solid dark energy ($w_X=1/3$), i.e., 
$\dot{x}=y$,
$\dot{y}=-\frac{1}{2} \Omega_{\text{top},0} -3\Omega_{\omega,0} x^{-7}
+ \Omega_{\Lambda,0} x$ 
where $\Omega_{\text{top},0}$ is density parameter for solid dark energy.
The acceleration region is situated on the right from the saddle point. 
Therefore, the trajectories of recollapsing and then expanding models lie 
permanently in the acceleration domain. Algebraic curves, on which lie 
trajectories, are given in explicit form  from of first integral
$\frac{y^2}{2}=\frac{1}{2}\Omega_{k,0}+\frac{1}{2}\Omega_{\text{top},0}x+
\frac{1}{2}\Omega_{\omega,0}x^{-8}+\frac{1}{2}\Omega_{\Lambda,0}x^2$. 
The physical region is determined by the condition $x>x_{0} \colon V(x_{0})=0$ 
where $x_{0}$ is the critical point.}
\label{fig:3}
\end{figure}

\begin{figure}
\includegraphics[width=0.8\textwidth]{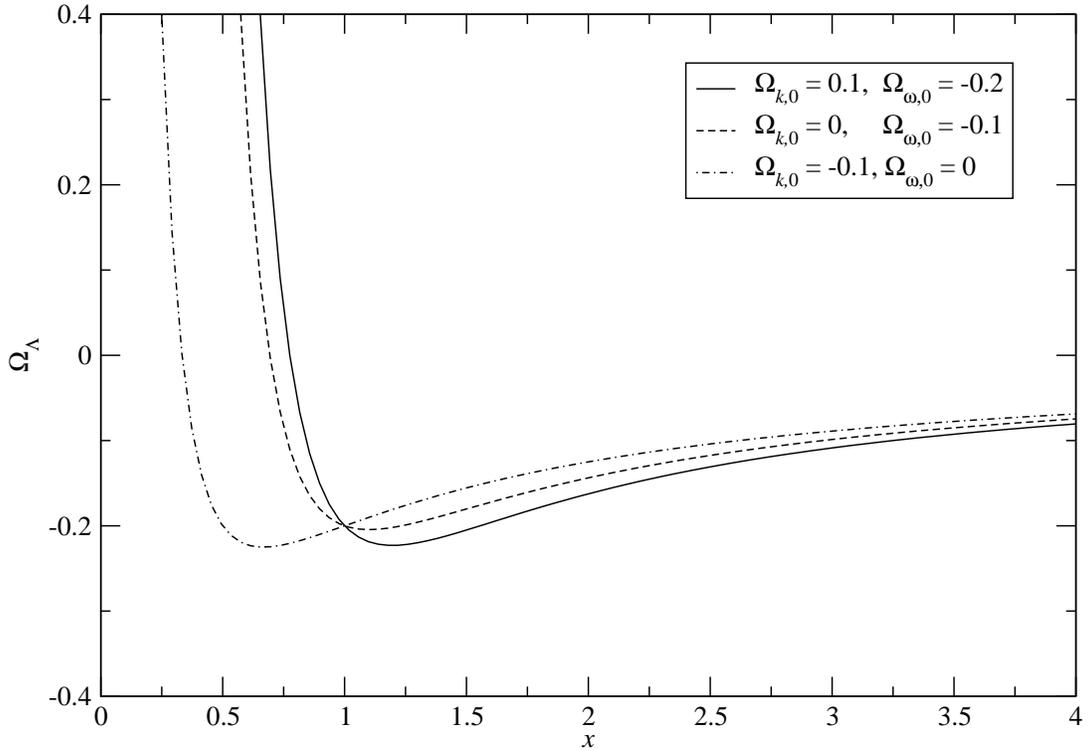}
\caption{
Diagram of the $\Omega_{\Lambda}(x)$ density parameter given by relation:
$\Omega_{\Lambda}(x) = - \frac{\Omega_{k,0} x^2 + \Omega_{\text{m},0} x^3
+ \Omega_{\omega,0}}{x^4}$
for different values of ($\Omega_{k,0}$, $\Omega_{\text{m},0}$,
$\Omega_{\omega,0}$). Forbidden (non physical) region: $y^2/2<0$ is situated
under the characteristic curve. The lines $\Omega_{\Lambda}(a)=\text{const}$ 
gives qualitative classification of possible path evolutions in the 
configuration space. The minimum is reached for $\Omega_{k,0}=0$ at
$x_{\text{min}}=\sqrt[3]{\frac{-4\Omega_{\omega,0}}{\Omega_{\text{m},0}}}
=\sqrt[3]{\frac{16\omega_0^2}{3\Omega_{\text{m},0}}}$. Note the existence of
oscillating models without the singularity for $\Omega_{\Lambda,0}<0$.}
\label{fig:1}
\end{figure}

\begin{figure}
\includegraphics[width=0.7\textwidth,angle=-90]{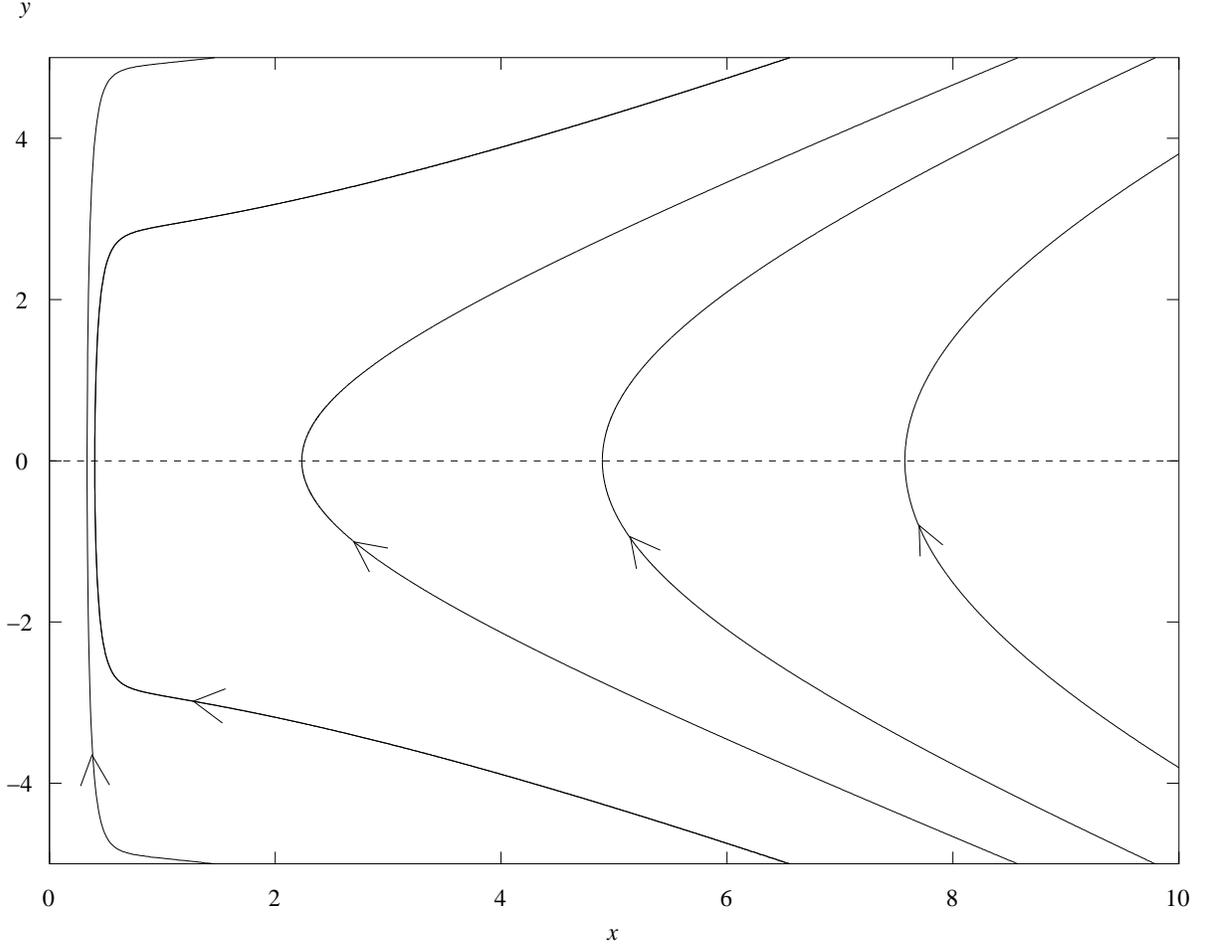}
\caption{
The phase portrait of the FRW model with global rotation, the cosmological 
constant and dust given by system (\ref{eq:54}). All trajectories start 
from state $x=\infty, \dot{x}=-\infty$ and after bouncing expand to infinity.
Cosmologies with rotation are represented only by
physical part of the phase curve starting from point $x=x_{\text{min}}, y=0$. 
There is no critical point in the physical domain.}
\label{fig:2}
\end{figure}

\begin{figure}
\includegraphics[width=0.7\textwidth,angle=-90]{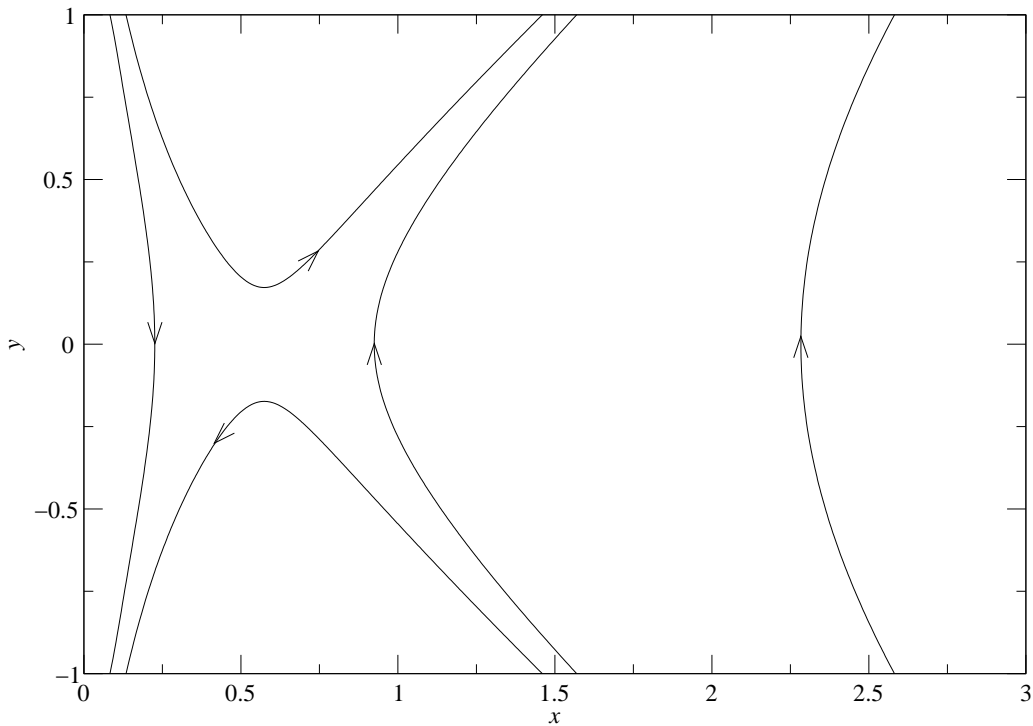}
\caption{The phase portrait of system (\ref{eq:54}) with the cosmological 
constant but without global rotation $\Omega_{\omega,0} = 0$. 
The acceleration region is situated on the left from the saddle point.
Therefore, the recollapsing and then expanding models lie
permanently in the acceleration domain. The Eddington model also belongs to
this region. The Lema\^itre-Eddington type models with characteristic loitering
phase start accelerating in the middle of quasi-static phase.}
\label{fig:4}
\end{figure}

For the FRW model with global rotation, the cosmological constant and 
dust we have
\[
w_{\omega} = \frac{10-6\gamma}{3}, \qquad w_{k}=\frac{2}{3}, \qquad 
\gamma = 1
\]
and
\begin{subequations}
\label{eq:54}
\begin{align}
\label{eq:54a}
\dot{x} &= y \\
\label{eq:54b}
\dot{y} &= - \frac{1}{2} \Omega_{\text{m},0} x^{-2} - \Omega_{\Lambda,0} x
- \Omega_{\omega,0} x^{-3}
\end{align}
\end{subequations}
System (\ref{eq:54}) has the first integral given by
\begin{equation}
\label{eq:55}
y^2 = - \frac{1}{2} \Omega_{\text{m},0} x^{-1} + \Omega_{k,0}
+ \Omega_{\Lambda,0} x^2 + \Omega_{\omega,0} x^{-2}.
\end{equation}
The classification of all admissible evolutional paths in the configuration
space is demonstrated in Fig.~\ref{fig:1}. The phase portrait which 
visualizes the all evolutions for all initial conditions is presented 
on Fig.~\ref{fig:2}. For comparison the phase portrait with the cosmological 
constant, and topological defects is in Fig.~\ref{fig:3}. 
The concordance $\Lambda$CDM model is shown on Fig.~\ref{fig:4}. 

The critical points of (\ref{eq:54}) are solutions of the equations
\begin{equation}
\label{eq:56}
V(x_0,y_0)=0 \Leftrightarrow \Omega_{\omega,0} x^{-2}
+ \Omega_{\text{m},0} x^{-1} + \Omega_{k,0} + \Omega_{\Lambda,0} x^2 = 0
\end{equation}
and
\begin{equation}
\label{eq:57}
\left. \frac{\partial V}{\partial x} \right|_{(x_0,y_0)}=0
\Leftrightarrow - \Omega_{\omega,0} x^{-3}
- \frac{1}{2} \Omega_{\text{m},0} x^{-2} + \Omega_{\Lambda,0} x = 0
\end{equation}
where $\sum_{i} \Omega_{i,0} = 1$.
 
Trajectories of the system belong to the domain admissible for motion
\[
\mathcal{D} = \{ (x,y) \in \mathbf{R}^2 \colon x_{\text{min}}<x
\ \text{and} \ V(x)\le 0 \}.
\]
Therefore, critical points lie on the boundary
$\partial \mathcal{D}= \{x \colon V(x)=0 \}$
\[
V(x) = - \frac{1}{2} x^{-2} (\Omega_{\omega,0} + \Omega_{\text{m},0} x
+ \Omega_{k,0} x^2 + \Omega_{\Lambda,0} x^4).
\]
 
To find the solution $x_0$ of (\ref{eq:56}) and (\ref{eq:57}) we consider 
the probe equation which helps us to formulate necessary condition for 
$x_0 \ne 0$ to be solution if only 
\[
- \Omega_{\omega,0} + \Omega_{k,0} x^2 + 3 \Omega_{\Lambda,0}x^4 = 0.
\]
Of course, if both $\Omega_{\Lambda,0}$ and $\Omega_{k,0}$ are positive 
then there is no critical point in the finite domain. 
 
Let us consider the case of $\Omega_{k,0} < 0$. Then
\[
x_0 = \sqrt{\frac{-\Omega_{k,0}
- \sqrt{\Omega_{k,0}^{2} + 12 \Omega_{\omega,0} \Omega_{\Lambda,0}}}
{6\Omega_{\Lambda,0}}}, \qquad y_0=0.
\]
where it is assumed that rotation is sufficiently small to ensure that 
real $x_0$ exists, i.e.,
\[
- \Omega_{\omega,0} < \frac{\Omega_{k,0}^{2}}{12 \Omega_{\Lambda,0}}.
\]
In the opposite case if the contribution coming from rotation is
sufficiently large, the domain admissible for motion is empty.
 
Finally, the smallest value of $x=x_{0}$ is given by  $V(x_{0})=0$ 
because the critical points must be always a zero of the potential function.

From the physical point of view the critical point of this type
represents the static Einstein universe.
 
Let us briefly comment now on the case of vanishing rotation in the above 
context. Then the domain admissible for motion is for positive $x$. 
In the phase space all critical points are situated at the $x$-axis, at the 
point
\[
x_0 = \sqrt{-\frac{\Omega_{k,0}}{3\Omega_{\Lambda,0}}}, \qquad
y_0 = 0.
\]
Therefore, we have always static critical points if only $\Omega_{k,0}$
and $\Omega_{\Lambda,0}$ are of opposite signs.
 
The character of the critical point is determined from eigenvalues of
the linearization matrix at critical point or equivalently from the
convectivity of $V(x)$
\begin{equation*}
\frac{\partial^2 V}{\partial x^2} = - 3\Omega_{\omega,0} x^{-4} 
- \Omega_{\text{m},0} x^{-3} - \Omega_{\Lambda,0}
\end{equation*}
but from (\ref{eq:56}) we have at the critical point
\[
- \Omega_{\omega,0} x_0^{-4} = \frac{1}{2} \Omega_{\text{m},0} x_0^{-3}
- \Omega_{\Lambda,0}.
\]
 
Then for any $x_0$ we have
\begin{align*}
\left. \frac{\partial^2 V}{\partial x^2} \right|_{(x_0,0)}
&= - \frac{5}{2} \Omega_{\text{m},0} x_{0}^{-3} - 4 \Omega_{\Lambda,0} < 0
\qquad \text{or in general} \\
\left. \frac{\partial^2 V}{\partial x^2} \right|_{(x_0,0)}
&= \det A = -\lambda_1 \lambda_2
\end{align*}
where $\lambda_i$ $(i=1,2)$ are eigenvalues of the linearization matrix $A$
of the system
\[
A = \left[ \begin{array}{cc}
0 & 1 \\
- \frac{\partial^2 V}{\partial x^2} & 0
\end{array} \right].
\]
Then eigenvalues are solutions of the characteristic equation
$\lambda^2 - (\tr A) \lambda + \det A = 0$ for the matrix $A$.
 
Finally, we show that in the physical domain our system with positive
curvature $k>0$ has at least one critical point which is represented on phase 
plane by the saddle point.
 
Let us consider the phase portrait of the system with global rotation and dust 
matter on the phase plane $(x,y)$. We assume the existence of the non-vanishing
cosmological term because the value $\Omega_{\Lambda,0} \ne 0$ is required
by consistency of the model with global rotations with SNIa data.
 
If we consider the standard FRW cosmology on the $(x,y)$ plane then situation 
is presented in Fig.~\ref{fig:4}. The flat model trajectory 
separates the region of the model with negative and positive curvature.
The critical point is a saddle and it represents the static Einstein
solution. Non-generic cases corresponding to separatices going in or out
of the saddle point. The acceleration region is situated on the right from
the saddle point. Therefore the recollapsing and then expanding models lie
permanently in the acceleration domain and there is a transition from 
decelerating to accelerating epochs. The Eddington model is also in
this region. The Lema{\^{\i}}tre-Eddington models start accelerating in the
middle of the quasi static phase. As we can see from Fig.~\ref{fig:3}
the global rotation contribution introduces qualitative changes of behaviour 
of trajectories, i.e., there is no homeomorphism which preserves directions
along trajectories and transforms the corresponding the trajectories. In our 
case there is no critical points on the phase plane at the finite domain 
(Fig.~\ref{fig:3}). It can be shown that in general if only
$a > \left( \frac{3\rho_0}{4\Lambda}\right)^{1/3} = a_{\text{min}}$, then
there is no critical point representing a static Einstein universe. 

\begin{figure}
\includegraphics[width=0.7\textwidth,angle=-90]{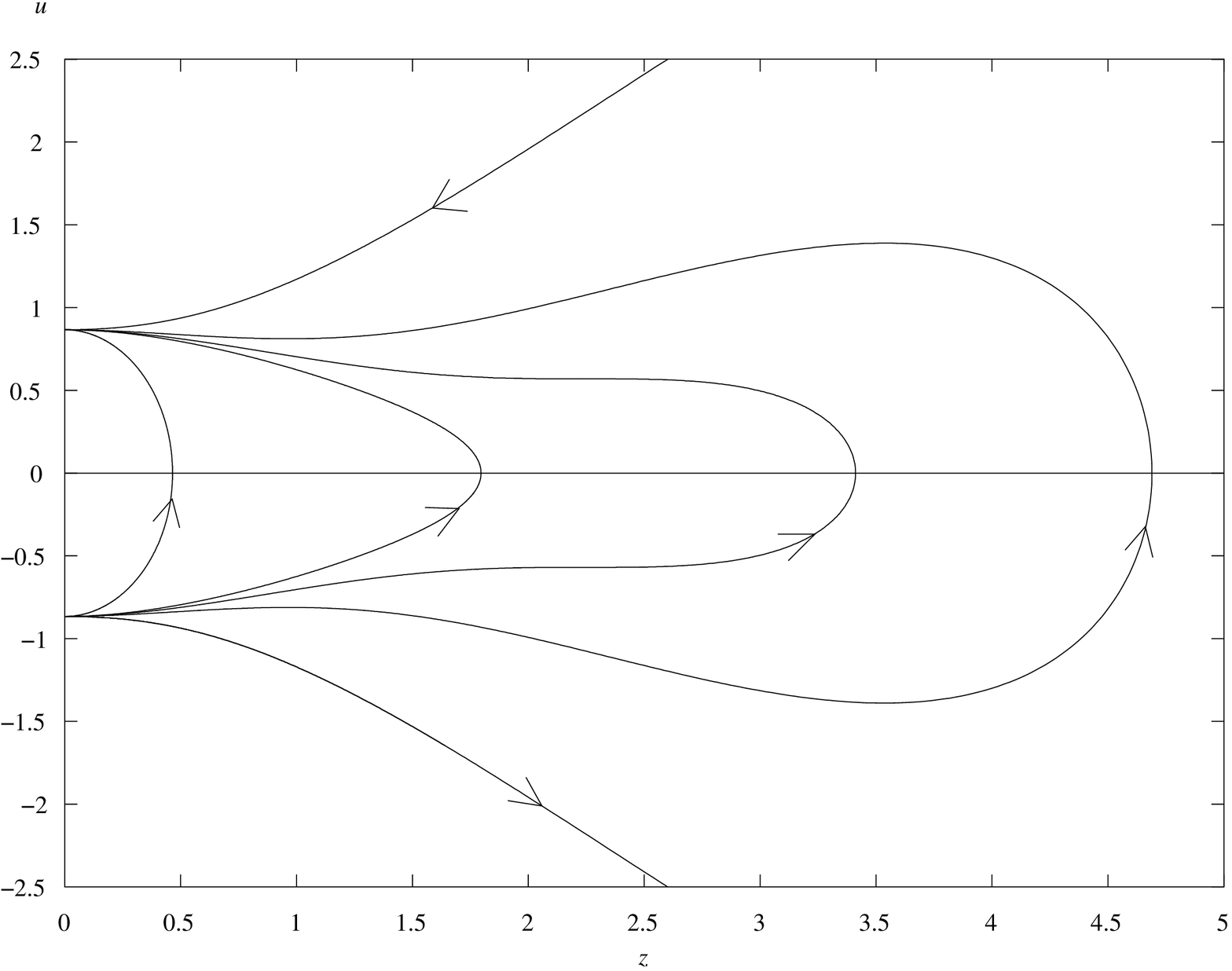}
\caption{The phase portrait of the model (\ref{eq:59}) 
with global rotation $\Omega_{\omega,0} \ne 0$, the cosmological constant 
and dust in the projective coordinates $(v,u) \colon v=1/x$, $u=y/x$. 
By introducing such coordinates we can observe their behaviour
at infinity. There are two types of trajectories which expand
to the maximum scale factor and then recollapse and vice versa.
In the first type of evolution the closed models start with the
singularity $a(0)=0$ and $\dot{a}=0$, then reach the maximum radius
and recollapse while the other class of models recollapse from
$a(0)=\infty$ and $\dot{a}=-\infty$ to the minimum value of a,
and then expand to infinity. In the generic case trajectories
start from the anti-de Sitter stage and land at the de Sitter state.}
\label{fig:5}
\end{figure}

To analyse system (\ref{eq:54}) at infinity it is useful to introduce the 
projective map $(v,u) \colon v=1/x$, $u=y/x$ on the plane and 
\begin{subequations}
\label{eq:59}
\begin{align}
\label{eq:59a}
\dot{v} &= - uv \\
\dot{u} &= -\frac{1}{2} \Omega_{\text{m},0} v^3 - \Omega_{\omega,0} v^4
+ \Omega_{\Lambda,0} - u^2
\end{align}
\end{subequations}
Then, circle at infinity is covered by line $v=0$ ($x= \infty$). The 
dynamical system with the non-zero cosmological
constant and non-vanishing global rotation in the finite region is presented in
Fig.~\ref{fig:5}. The region $x < x_{\text{min}}$ is forbidden because the 
global rotation contribution cannot dominate the matter contribution. In this 
case there is one type of trajectories which recollapses from the initial 
singularity and then expands after reaching the minimum size of scale factor. 
Hence the expanding de Sitter model is a global attractor and the contracting 
de Sitter one is a global repeller (Fig.~\ref{fig:5}). All point
at infinity which lies on $v=0$ circle are hyperbolic, therefore the system
is structurally stable.

\section{The flatness and lambda problems}
 
If we choose the standard FRW cosmology without the global
rotation then the conservation equation for perfect fluid (\ref{eq:3}),
satisfying the equation of state $p_i=(\gamma_i-1)\rho_i$, gives 
$\rho_i \propto a^{-3\gamma_i}$ and the $\rho/3$ term, in the Friedmann first 
integral (called also the Friedmann equation). Let us consider first these
problems by setting $\Lambda=0$,
\begin{equation}
\label{eq:61}
\frac{\dot{a}^2}{a^2} = -2V(a) = \frac{\rho_0}{3} a^{-3\gamma} + \frac{2}{3}
\frac{\omega_{0}^{2} a^{6\gamma -10}}{(3\gamma-4)} + \frac{k}{a^2}
\end{equation}
where $\sigma_0=0$, i.e., shear vanishes.
 
The matter term dominates the curvature term $ka^{-2}$ at large as long as 
the matter stress obeys $\rho + 3p < 0$, $\rho + p \ge 0$, that is if 
$0 \le \gamma <2/3$. This is what we describe as the flatness problem. 
Since the scale factor then evolves as $a(t) \propto t^{2/3\gamma}$ if 
$\gamma>0$ or if $\gamma = 1$ as $\exp(H_{0}t)$. We can observe that it grows 
faster than the proper size of the particle horizon scale ($\propto t$) so 
long as $\gamma<2/3$. Therefore, a sufficiently long period of evolution 
during which expansion is dominated by matter with negative pressure with 
$\rho + 3p<0$ can solve the flatness and horizon problems. Such a period of 
accelerated expansion is called inflation \cite{Barrow99b}. However, as 
Barrow noted there is a flatness problem if the cosmological constant is 
added to the right hand side of the Friedmann equation. Then, to explain why 
it does not dominate matter term $\rho/3$ (at large $a(t)$) it is assumed 
that there exists the period in the early evolution of the universe during 
which weak energy condition for the matter is broken, i.e, $\rho +p<0$. 
This is what Barrow called the cosmological constant problem. For the 
standard FRW model the existence of corresponding $\gamma < 0$ from equations 
of state mean that $\dot{a} < 0$ which is the case of phantoms. Therefore, 
models with global rotation provide a solution of the flatness
problem. It is formally equivalent to presence in the model some fictitious
additional noninteracting fluid which obeys the equation of state
$p_{\omega}=\frac{7-6\gamma}{3} \rho_{\omega}$,
$\rho_{\omega}=\rho_{\omega_0}\left(\frac{a}{a_0}\right)^{6\gamma-10}$,
$\rho_{\omega_0}=\frac{2}{3\gamma-4}\omega_{0}^{2}$. Then both weak and 
strong energy conditions for pure ``rotational fluid'' means
\begin{gather*}
\rho_{\omega} + p_{\omega} = w_{\omega} \rho_{\omega}
= \frac{10-6\gamma}{3} \rho_{\omega}  \ge 0 \\
\rho_{\omega} + 3p_{\omega} = 2\rho_{\omega}(4-3\gamma) \ge 0.
\end{gather*}
In the special case of dust matter we have inequality if the above conditions 
are violated 
\begin{align*}
&\rho_{\omega} + p_{\omega} = \frac{4}{3} \rho_{\omega} < 0, &
&\rho_{\omega}=\rho_{\omega_0}x^{-4} \\
&\rho_{\omega} + 3p_{\omega} = 2\rho_{\omega} < 0, &
&\rho_{\omega_0} = -2 \omega_{0}^{2},
\end{align*}
i.e., both energy conditions are violated.
 
Finally, the strong and weak energy conditions for density energy of dust and 
rotational fluids requires respectively
\begin{gather*}
(\rho_{\text{m}}+\rho_{\omega}) + 3p_{\omega} = \rho_{\text{m},0} a^{-3}
\left( 1 - \frac{2}{\rho_{\text{m},0}} \omega_{0}^{2} a^{-1}\right) \ge 0, \\
(\rho_{\text{m}}+\rho_{\omega}) + p_{\omega} = \rho_{\text{m},0} a^{-3}
\left( 1 - \frac{8}{3\rho_{\text{m},0}} \omega_{0}^{2} a^{-1}\right) \ge 0.
\end{gather*}
 
Thus, during the evolution, first, the strong energy condition is violated for
$a < \frac{2}{\rho_{\text{m},0}} \omega_{0}^{2} = 
\frac{2}{3H^2_0} \frac{1}{\Omega_{\text{m},0}}$
and the weak one is violated for $a < \frac{8}{3\rho_{\text{m},0}}$. We can also 
find the interval ${2\rho_0} < a < \frac{8}{3\rho_{\text{m},0}}\omega_{0}^{2}$ 
in which the strong energy condition is violated, whereas weak energy 
condition is still satisfied. It is possible because $\rho_{\omega}$ is 
negative. These are sufficient conditions to solve the flatness problem.

Because of the existence of the period in which $\rho+p<0$ (phantom), the 
cosmological constant problem can be solved and $\dot {a}<0$ is an unnecessary 
requirement. Unfortunately, the future horizon problem cannot be solved and 
the corresponding condition $\gamma \le 2/3$ is identical to that of the
solution of the horizon (or flatness) problem in the case with vanishing 
rotation.

Let us note that if we consider the case of the non-zero $\Lambda$ term,
that is contributed by a pressure $p_{\Lambda}=-\rho_{\Lambda}$ then
$\Lambda$ term on the right-hand side of (\ref{eq:61}) falls off faster
than the curvature and matter density terms so long as $\gamma < 2/3$
and $\gamma<0$ respectively. The contribution term from the global
rotation falls off faster then curvature $\Lambda$ and matter
density term so long as $\gamma<4/3$ (radiation), $\gamma<5/3$,
$\gamma<10/9$, respectively.
 
Therefore, the cosmological constant term $\Lambda$, curvature term is
proportional to $a^{-2}$ and term connected with global rotation falls off
faster than $8\pi G\rho/3$ term in the FRW equation at large $a(t)$ if
respectively $\gamma<0$ and $\gamma<2/3$ and $\gamma<10/9$, i.e., $\gamma<0$.

\section{Homologous universe and its angular momentum}

In this section we demonstrate how Wesson's argument can be generalised and
how it works in our case \cite{Wesson79,Wesson83}. Wesson argued that 
because the dynamics of a gravitationally bound rotating system admits 
self similarity group of symmetry the relation between angular momentum $J$ 
of a astronomical system and its mass $J \sim M^2$ can be constructed 
due to constancy of the invariants of this kind of symmetry.
In this section we demonstrate how Wesson's argument can be generalised and
how it works in our case. We generalise Wesson's argumentation to the case
of universe dynamics by consideration of Lie (continuous) symmetries of
the FRW models with rotation. The Lie group theory of symmetries analysis
gives us information about the group of symmetry transformations admissible 
by the differential equation structure. They preserve the structure of basic 
dynamical equations. In the set of solutions, the action of an admissible 
group includes a certain algebraic structure which can be used to find 
a family of new solutions from the known ones 
\cite{Barenblatt96,Bluman89,Stephani89,Ovsiannikov82}.

As is well known \cite{Kippenhahn80}, the new solution for the system
equation, describing the static stars (in Newtonian as well as in General
Relativity), can be obtained from the known ones through the homologous
transformation \cite{Collins77,Biesiada88,Biesiada88b}. We can find the 
homologous transformation of the symmetry type (i.e., preserving structure of
differential equations) for the considered dynamics of the FRW model with 
rotation.
It would be useful to rewrite basic dynamical equations (\ref{eq:9}) and
(\ref{eq:3}) to the form in which instead of ($H$, $\rho$, $\omega$) variables
we have ($a$, $\rho$, $\omega$). Then we obtain
\begin{subequations}
\label{eq:149}
\begin{align}
\frac{da}{dt} &= \sqrt{\frac{\rho a^2}{3} -k - \frac{2}{3}\omega^2 a^2} \\
\frac{d\rho}{dt} &= - 2\gamma \rho
\sqrt{\frac{\rho a^2}{3} -k - \frac{2}{3}\omega^2 a^2} a^{-1} \\
\frac{d\omega}{dt} &= - \omega (5 - 3\gamma)
\sqrt{\frac{\rho a^2}{3} -k - \frac{2}{3}\omega^2 a^2} a^{-1}.
\end{align}
\end{subequations}
Now let us consider the differential equations system
\[
\frac{du^i}{dx} = f^{i}(x, u^1, \ldots, u^m),\quad i=1, \ldots, m,
\]
and here the space of independent variable is denoted as $x$, dependent 
one $u^{\alpha}$ and its first derivatives, say 
$u^{\alpha'} (\alpha = 1, \ldots, m)$.
The action of the Lie group $G$ of point transformation in the space
$(x,u)$ is described by an infinitesimal operator $\mathcal{X}$---a generator
of the symmetry
\[
\mathcal{X} = \zeta(x,u^1, \ldots, u^m)
\frac{\partial \phantom{x}}{\partial x}
+ \sum\limits_{i=1}^m
\eta^i(x,u^1, \ldots, u^m) \frac{\partial \phantom{u^i}}{\partial u^i}.
\]
The point transformation generated by $\mathcal{X}$ is called homologous if
$\zeta= ax$ and $\eta^i = g^i u^i$, where $a, g^i(i=1,\ldots,m)$ are constants.
Of course, if $\mathcal{X} \equiv \lambda^s(u) \partial_s$, then the finite
transformation of symmetry $u \to \bar{u}$ are given as a solution of
equation $\frac{d\bar{u}^s}{dt} = \lambda^s(u)$ with the initial conditions
$\bar{u}^s (\tau=0)= u^s$ ($s=1, \ldots,m$). The action of the Lie group
transformation can be extended from the ($x,u$) space to the ($x,u,u'$) space,
i.e for the first derivatives. On the other hand, $s$-th order (ordinary or
partial) differential equation
\[
F(x,u(x),u^1(x), \ldots, u^s(x)) = 0
\]
defines a certain manifold $\mathcal{M}$ in the space ($x, u, u^1, \dots, u^s$).
We say that $F$ is invariant with respect to the action group $G$, provided
that manifold $\mathcal{M}$ is a fixed point with respect to the $s$-th
extension of $G$, i.e., $G^s(\mathcal{M})= \mathcal{M}$. In the terms of
the infinitesimal operator it means that
\[
\mathcal{X}^s F \vert_{F=0} =0
\]
where $\mathcal{X}^s$ is the extented operator on the first derivatives
\begin{gather*}
\mathcal{X}^s = \mathcal{X} + \zeta_i^{\alpha}(x,u,u')
\frac{\partial \phantom{u^i}}{\partial u_i^{\alpha}}, \qquad
u_i^{\alpha} = \frac{\partial u^{\alpha}}{\partial x^i}, \\
\zeta_i^{\alpha} = D_i(\eta^{\alpha}) - u_j^{\alpha} D_j(\zeta^j), \qquad
D_i=\frac{\partial \phantom{x^i}}{\partial x^i} + u_i^{\alpha}
\frac{\partial \phantom{u^a}}{\partial u^{\alpha}}.
\end{gather*}
The prolonged operators, which are generators of symmetry of equations in
the space $(x,u)$, form the structure of a Lie algebra of the fundamental
group.
 
The Lie method adopted to our case gives us the symmetry operator in the form
\begin{equation}
\label{eq:150}
\mathcal{X} = \alpha \rho \frac{\partial}{\partial \rho}
- \frac{\alpha}{2} a \frac{\partial}{\partial a}
+ \frac{\alpha}{2} \omega \frac{\partial}{\partial \omega}
- \frac{\alpha}{2} t \frac{\partial}{\partial t},
\end{equation}
where $\alpha = \text{const}$, i.e., the most general form of symmetries
constitutes a homologous transformation.
For our further analysis it is useful to introduce the notion of
invariants of symmetry operator $\mathcal{X}$. For the infinitesimal operator
$\mathcal{X}$ there is $m$ independent invariants which are solutions of
the following system
\begin{equation}
\label{eq:151}
\frac{dx}{\zeta(x,u^1, \ldots, u^m)}
= \frac{du^1}{\eta^1(x,u^1, \ldots, u^m)} = \cdots
= \frac{du^m}{\eta^m(x,u^1, \ldots, u^m)}.
\end{equation}
In our case we can recover two invariants in the form
\begin{equation}
\label{eq:152}
J_1 = \rho a^2, \qquad J_2 = \omega a.
\end{equation}
They exactly correspond to Wesson's invariants expressed in dimensionless form
$\eta_1 \equiv G\rho a^2 /c^2$, $\eta_2\equiv \omega a/c$. The significance
of the obtained invariant homologous transformation of (\ref{eq:34}) is the
following.

1) They can be used as good new variables to integrate the system by
lowering its dimension. Then we obtain
\begin{equation}
\label{eq:153}
\frac{dJ_1}{d\eta} = (2 - 3\gamma)J_1, \qquad
\frac{dJ_2}{d\eta} = (3\gamma - 4)J_2, \qquad
\eta \colon Hdt = d\eta.
\end{equation}

2) The invariant can be used to obtain a new solution from the known ones
using the homologous theorem. If $a(t)$ is a solution of (\ref{eq:34}),
then because
$a(t)t^{-1}$ is invariant, $a(e^{-\alpha \tau / 2} t)(e^{-\alpha \tau / 2}
t)^{-1}=\text{const}$ is also a solution, where the finite transformations
are given in the form
\begin{equation}
\label{eq:154}
\rho \to \bar{\rho} = \rho e^{\alpha \tau}, \quad
a \to \bar{a} = a e^{-\alpha \tau / 2}, \quad
\omega \to \bar{\omega} = \omega e^{\alpha \tau / 2}, \quad
t \to \bar{t} =  e^{-\alpha \tau / 2}
\end{equation}
where $\tau$ is a group parameter and $\alpha = \text{const}$.
Let us note that the finite transformation (\ref{eq:154}) preserves 
the angular momentum form
\begin{equation}
\label{eq:155}
J=\rho a^5 \omega = \bar{\rho} \bar{a}^5 \bar{\omega},
\end{equation}
i.e., this relation is invariant with respect to the homologous (or self
similarity) transformation group given by (\ref{eq:154}).
Physically the existence of this type of symmetry always means that a basic 
dynamical equation can be formulated in a dimensionless form.
 
It is interesting that (\ref{eq:155}) can be expressed in terms of
invariants by
\begin{equation}
\label{eq:156}
J = \frac{J_2}{J_1} M^2,
\end{equation}
and astrophysical relation $J \sim M^2$ which is closely obeyed over mass
range $10^{18}g$ to $10^{48}g$ can be extended to the Universe, together
with the Wesson's philosophy that self similarity transformation plays an
important role in astrophysics. This notion was introduced in astrophysics
by Rudzki \cite{Rudzki02}. We note that it can play an important role
in cosmology (see also \cite{Stephani89}).
 
Moreover the ratio $J_2/J_1$ in the equation (\ref{eq:156}) is constant in
the distinguished case of dust matter ($\gamma = 1$). Therefore, there is no
need for additional arguments which usually are taken from the astronomical
observation about the constancy $J_2/J_1$ as in the Wesson argumentation
because after integration of (\ref{eq:153}) we obtain immediately 
\begin{equation}
\label{eq:157}
\frac{J_2}{J_1} \sim e^{6(1 - \gamma) \eta},
\end{equation}
and only for $\gamma=1$ (dust) we obtain
\begin{equation}
\label{eq:158}
\frac{J_2}{J_1} = \mbox {const}.
\end{equation}

\section{Conclusion}

In this paper we studied the dynamics of the FRW universe with global rotation.
Our approach is simplest in that we formulate the dynamical problem in 
two-dimensional phase space. Moreover, we find the Hamiltonian formulation of 
the dynamics. Such visualisation has a great advantage because it allows us 
to analyse the acceleration problem in a clear way. Due to the existence of 
the Hamiltonian constraint, it is possible to make the classification of the 
qualitative evolution paths by analyzing the characteristic curve which 
represents the boundary equation in configuration space. Dynamical system 
methods gives us all evolutional paths for all possible initial conditions.
On the other hand, the representation of dynamics as a one dimensional
Hamiltonian flow allows us to make the classification of the possible evolution
paths in the configuration space. When we consider the dynamics of the FRW 
models with global rotation, a two-dimensional dynamical system in general, 
then there is a simple test of structural stability of
the system (a dynamical system $S$ is said to be structurally stable if
dynamical systems in the space of all dynamical system which are close to $S$
are topologically equivalent (physically realistic models on the plane should
be structurally stable).
Namely, if the right-hand sides of dynamical systems are in polynomial form,
the global phase portraits are structurally stable on $S^2$ ($R^2$ with adjoint
a Poincar{\'e} sphere) if, and only if, the number of critical points and limit
cycles is finite, each point is hyperbolic and there are no trajectories
connecting saddle points. Our conclusion is that the FRW models with global
rotation are structurally stable, i.e., following Peixoto's theorem dynamical
systems with rotation form open and dense subsets in the space of all
dynamical system on the plane. From the point of view of science modelling 
they are better and more adequate description of the observed universe than 
the FRW model without the cosmological term.

Our formalism gives natural base to express dynamical equations in the form of
the FRW model with some additional fictitious noninteracting multifluid which 
mimics the effects of global rotation. It satisfies in the generic case the
equation of state for radiation with some unusual negative energy density.
We find basic dynamical equations
describing the dynamics in the form of two-dimensional Hamiltonian dynamical
systems in which coefficients are just dimensionless observational density
parameters $\Omega_{i,0}$.
 
We also showed that global rotation produces a dark energy component but the
cosmological term is still required to explain SNIa data. We also 
demonstrates how classical cosmological problems can be solved due to the 
existence of global rotation. We also derive the observationally suggested 
relation between the angular momentum and mass of the object $J \propto M^2$ 
from the property of self similarity of dynamics of the FRW cosmology with 
rotation.

The presented formalism forms the base for our further analysis of 
the magnitude redshift relation and finding the best fitting for 
$\Omega_{\omega,0}$ from recently available SNIa data and X-ray gas 
measurements of mass fraction in clusters of galaxies or measurements of 
angular size of radiogalaxies.

\acknowledgments
M. Szyd{\l}owski acknowledges the support of the KBN grant no. 2 P03D 003 26.

\end{document}